\documentclass[aps,pra,twocolumn,groupedaddress]{revtex4-2}
\usepackage{graphicx}% Include figure files
\usepackage{dcolumn}% Align table columns on decimal point
\usepackage{bm}% bold math
\usepackage{mathtools}
\usepackage{physics}
\usepackage{hyperref,verbatim}
\usepackage{color}
\usepackage[commandnameprefix=always]{changes} % chadded, chdeleted, chreplaced

\begin{document}

%\preprint{}

\title{QED corrections of orders $m\alpha^6$ and $m\alpha^6(m/M)$ for HD$^+$ spin-independent rovibrational transitions beyond Born-Oppenheimer approximation}


\author{Zhen-Xiang Zhong}
\email{zxzhong@hainanu.edu.cn}
%\email{zxzhong@apm.ac.cn}
%\thanks{}
\affiliation{Center for Theoretical Physics, School of Physics and Optoelectronic Engineering, Hainan University, Haikou 570228, China}

\author{Ping Yang}
\affiliation{Innovation Academy for Precision Measurement Science and Technology, Chinese Academy of Sciences, Wuhan 430071, China}
\affiliation{University of Chinese Academy of Sciences, 100049, Beijing, China}
\author{Vladimir I. Korobov}
\affiliation{Bogoliubov Laboratory of Theoretical Physics, Joint Institute for Nuclear Research, Dubna 141980, Russia}

\author{Chun Li}
\affiliation{School of Mathematics, Nanjing University, Nanjing Jiangsu 210093, China}

%\author{Zong-Chao Yan}
%\affiliation{State Key Laboratory of Magnetic Resonance and Atomic and Molecular Physics, Innovation Academy for Precision Measurement Science and Technology, Chinese Academy of Sciences, Wuhan 430071, China}
%\affiliation{Department of Physics, University of New Brunswick, Fredericton, New Brunswick, Canada E3B 5A3}

\author{Ting-Yun Shi}
\affiliation{Innovation Academy for Precision Measurement Science and Technology, Chinese Academy of Sciences, Wuhan 430071, China}


\date{\today}

\begin{abstract}
The effective Hamiltonian of $m\alpha^6$ and $m\alpha^6(m/M)$ order corrections for hydrogen molecular ions has been derived in [ Z.-X. Zhong, \emph{et al.}, Phys. Rev. A {\bf98}, 032502(2018).], in this work we express the energy correction in the form of finite-value effective operators.
The cut-off regularization scheme is used to determine finite part of divergent operators of the leading-order recoil corrections.
Numerical calculations of first-order contributions are performed in the Hylleraas basis set.
Combining the second-order terms calculated in recent work [V. I. Korobov, \emph{et al.}, Mol. Phys. e2563023 (2025).], the $m\alpha^6$-order corrections for the fundamental rovibrational transition are obtained with an uncertainty three times smaller than in previous calculations.
\end{abstract}

%\keywords{}

\maketitle

\section{introduction}

{The spectroscopy of HD$^+$ (a hydrogen molecular ion composed of a proton, a deuteron, and one electron) has enabled determinations of the proton-to-electron mass ratio \cite{Koelemeij2007,Koelemeij2012,Biesheuvel2016,Alighanbari2018} and  relative fundamental constants \cite{Alighanbari2020,Patra2020}, with the ion trapped in the Lamb-Dicke regime and spectroscopic precision reaching $10^{-11}$ or better.}
The corresponding theoretical calculations {have spanned} two decades, started from the non-relativistic energies \cite{Korobov2000,Yan2003,Yan2007} and leading-order relativistic and radiative corrections \cite{Korobov2004hde,Korobov2006,Korobov2006HMI,Zhong2012,Korobov2012BL,Zhong2013BL}, followed by $m\alpha^6$ order correction based on the Born-Oppenheimer (BO) approximation \cite{Korobov2007BO,Korobov2008HMI} and $m\alpha^7$ order corrections \cite{Korobov2013,Korobov2014PRA,Korobov2014PRL}.
In 2017, the theoretical spectroscopy reached a precision of $10^{-12}$ level by including adiabatic correction to the $m\alpha^6$ order contribution and the main contribution of $m\alpha^8$ order correction \cite{Korobov2017ppt}.
An update result adopting fundamental constants from CODATA 2018~\cite{CODATA2018} was presented in 2021 \cite{Korobov2021}.

In the review of all the calculations, the non-relativistic energies and leading-order relativistic and radiative corrections has been independently verified, while corrections of $m\alpha^6$ order and higher performed by only one group.
In 2018, Zhong and coworkers derived spin-averaged effective Hamiltonian of orders $m\alpha^6$ and $m\alpha^6(m/M)$ for hydrogen molecular ions \cite{Zhong2018NRQED}, which made an independent verification to HD$^+$ $m\alpha^6$ order calculations \cite{Korobov2007BO,Korobov2008HMI,Korobov2017ppt} possible.
It should be noted that the the recoil part contained some inaccuracies in derivation of {relativistic} correction in Ref. \cite{Zhong2018NRQED}, which was corrected recently in \cite{Korobov2025reg,Korobov2025rec}.
In this work, we intend to calculate the first-order contribution of orders $m\alpha^6$ and $m\alpha^6(m/M)$ to relativistic corrections beyond the BO approximation.
Thus the effective Hamiltonians are treated in the complete three-body formalism. Our approach accounts for nuclear mass dependence through the first two terms of the $(m/M)$ expansion, omitting higher-order recoil corrections. The second-order contributions were recently calculated in \cite{Korobov2025sec} and we will use them to get the complete  $m\alpha^6$ order correction to the spin-averaged bound-state energies in the HD$^+$ ion.

This article is organized as follows: section II presents effective Hamiltonian, section III shows procedure of numerical calculations for the first-order and second-order perturbation contributions, numerical results and a conclusion are provided in section IV.
Atomic units ($\hbar=m_e=e=1$) are used throughout this work, unless otherwise stated.
Physical constants of CODATA~2018 \cite{CODATA2018} are adopted.

\section{the $m\alpha^6$ order effective Hamiltonian}

In NRQED theory, an energy level of a light atomic system can be expanded in powers of the fine structure constant $\alpha$:
\begin{align}
	E(\alpha) = E^{(2)} + E^{(4)} + E^{(5)} + E^{(6)} + E^{(7)} + O(\alpha^8),
	\label{eq:energy-expansion-alpha}
\end{align}
where $E^{(n)}$ represents a contribution of order $m\alpha^n$ and may include nuclear recoil terms.
Each term $E^{(n)}$ can be obtained from the expectation value of the corresponding effective Hamiltonian.

In Eq.~\eqref{eq:energy-expansion-alpha}, $E^{(2)} \equiv E_0$ is the eigenvalue of the nonrelativistic Hamiltonian $H^{(2)} \equiv H_0$ with the associated eigenstate $\psi_0$ satisfying $H_0\psi_0 = E_0\psi_0$, where
\begin{align}\label{eq:H0}
	H_0 = \frac{\vb{p}_e^2}{2m_e} + \frac{\vb{p}_1^2}{2m_1} + \frac{\vb{p}_2^2}{2m_2}
	+ \frac{z_e z_1}{r_1} + \frac{z_e z_2}{r_2} + \frac{z_1 z_2}{r_{12}}.
\end{align}
The Hamiltonian $H_0$ is expressed in the center-of-mass frame with the constraint $\vb{p}_1 + \vb{p}_2 + \vb{p}_e \equiv 0$.
The relative position vectors between particles are defined as
\begin{align}
	\vb{r}_1 = \vb{r}_e - \vb{R}_1,\quad
	\vb{r}_2 = \vb{r}_e - \vb{R}_2,\quad
	\vb{r}_{12} = \vb{R}_2 - \vb{R}_1,
\end{align}
where $\vb{r}_e$, $\vb{R}_1$, and $\vb{R}_2$ denote the position vectors of the electron and nuclei 1 and 2, respectively.
For convenient, the Coulomb potential is expressed in terms of  pairwise interactions
\begin{align}
	V=V_1+V_2+V_{12}\,,
\end{align}
where $V_i=z_ez_i/r_i$ and $V_{12}=z_1z_2/r_{12}$.

The term $E^{(4)}$ in Eq.~\eqref{eq:energy-expansion-alpha} corresponds to the expectation value of the total Breit-Pauli Hamiltonian $H^{(4)}$.
The second-order perturbation of $H^{(4)}$ contributes to the spin-averaged $m\alpha^6$-order correction.
Since we are interested in the nonrecoil and leading-order recoil corrections of $H^{(6)}$, we consider only the required Hamiltonians:
\begin{align}
	H^{(4)} &= H_B + H_R + H_S,\\
	H_B &= -\frac{\vb{p}_e^4}{8m_e^3} - \frac{\rho_1+\rho_2}{8m_e^2}, \label{eq:ma4-HB}\\
	H_R &= \sum_a -\frac{z_e z_a}{m_e m_a} p_e^i \mathcal{W}^{ij}_a p_a^j, \label{eq:ma4-HR}\\
	H_S &= \sum_a -z_e z_a \bigg[\frac{1 + 2a_e}{2m_e^2} \frac{1}{r_a^3} (\vb{r}_a \times \vb{p}_e)
	\nonumber\\&\displaystyle\hspace{20mm}
	- \frac{1 + a_e}{m_e m_a} \frac{1}{r_a} (\vb{r}_a \times \vb{p}_a)\bigg] \cdot \vb{s}_e,
	\label{eq:ma4-HS}
\end{align}
where $a$ runs over the two nuclei, $\rho_a=-\Delta V_a=4\pi z_ez_a\delta(\vec{r}_a)$, $\mathcal{W}^{ij}_a=(\delta^{ij}+r_a^ir_a^j/r_a^2)/(2r_a)$, and $a_e$ represents the electron's anomalous magnetic moment.
%The leading-order radiative correction $E^{(5)}$ is not considered in this work.

The focus of this paper is on the $m\alpha^6$ order correction, denoted as $E^{(6)}$.
This energy correction consists of several contributions of both order $m\alpha^6$ and order $m\alpha^6(m/M)$.
The first set of contributions arises from the low-energy NRQED Hamiltonian~\cite{Zhong2018NRQED} :
\begin{align}
	H_\text{rel}^{(6)} &= \sum_{i=1}^6 \delta H_i, \label{eq:H6_rel} \\
	\delta H_1 &= \frac{\vb{p}_e^6}{16m_e^5}, \label{eq:dH1}\\
	\delta H_2 &= \frac{1}{128m_e^4} [\vb{p}_e^2, [\vb{p}_e^2, V]]
	+ \frac{3}{64m_e^4} \left\{\vb{p}_e^2, \rho_1+\rho_2\right\}, \label{eq:dH2}\\
	\delta H_3 &= \sum_a -\frac{1}{32m_e^3 m_a} [\vb{p}_a^2, [\vb{p}_e^2, V_a]], \label{eq:dH3}\\
	\delta H_4 &= -\frac{1}{8m_e^2} \{\vb{p}_e^2, 2H_R\}, \label{eq:dH4}\\
	\delta H_5 &= \sum_a \frac{1}{2m_a}
	\bigg\{
	\left[\frac{z_ez_a}{2r_a} \left(\delta^{ij} + \frac{r_a^i r_a^j}{r_a^2}\right) \frac{p_e^j}{m_e}\right]
	\nonumber\\&\times
	\left[\frac{z_ez_a}{2r_a} \left(\delta^{il} + \frac{r_a^i r_a^l}{r_a^2}\right) \frac{p_e^l}{m_e}\right]
	+ \frac{1}{2m_e^2} \bm{\varepsilon}_a^2
	\bigg\}, \label{eq:dH5}\\
	\delta H_6 &= \sum_a -\frac{1}{m_e m_a}
	\bigg\{
	[p_e^i, V] \mathcal{X}^{ij}_a [V, p_a^j]
	\nonumber
	\\&\displaystyle\hspace{20mm}
	+ p_e^i [\mathcal{X}^{ij}_a, \frac{\vb{p}_e^2}{2m_e}] [V, p_a^j]
	\bigg\},\label{eq:dH6}
\end{align}
where $\bm{\varepsilon}_a=-\bm{\nabla}V_a$ and $\mathcal{X}^{ij}_a = \left[r^i_a r^j_a - 3\delta^{ij} r^2_a\right]/(8r_a)$.
Curly brackets denote $\{A,B\}=AB+B^*A^*$.
Another contribution --- the pure recoil correction --- is determined by the high-energy region as the contact type interaction of the NRQED~\cite{Pachucki1995Pure}:
\begin{align}
	H_H^{(6)} = \sum_a \left(4\ln 2 - \frac{7}{2}\right) \frac{z_a^3}{m_a} \pi \delta(\vb{r}_a).
	\label{eq:H_H}
\end{align}
Formally, the contribution of the three-photon exchange diagrams in (\ref{eq:H_H}) diverges, but, as it was shown \cite{Pachucki1997ma6}, it is cancelled out by the corresponding term from the low-energy region.

The second set of contributions to $E^{(6)}$ arises from radiative corrections of orders $m\alpha^6$ and $m\alpha^6(m/M)$, denoted by $H_\text{rad}^{(6)}$ and $H_\text{rad-rec}^{(6)}$, respectively.
These corrections are derived within the bound-state QED.
The $m\alpha^6$-order nonrecoil contribution includes both one-loop and two-loop corrections and may be written in the form:~\cite{THBS1990,Eides2007}
\begin{align}
\begin{array}{@{}l}\displaystyle
	H_\text{rad}^{(6)} = {\sum_a}\frac{4\pi}{m_e^2}
	\bigg\{
	\left(\frac{139}{128} - \frac{\ln 2}{2} + \frac{5}{192}\right) z_a^2 \delta(\vb{r}_a)
\\[3mm]\displaystyle\hspace{10mm}
	- \frac{1}{4\pi^2} \left[\frac{2179}{648} + \frac{10}{27}\pi^2 - \frac{3}{2}\pi^2 \ln 2 + \frac{9}{4}\zeta(3)\right] z_a \delta(\vb{r}_a)
	\bigg\}.
	\label{eq:H_rad}
\end{array}
\end{align}
The radiative recoil high-energy contribution of the order of $m\alpha^6(m/M)$ is determined by the following expression~\cite{Eides2007,Blokland2002}:
\begin{align}
	H_\text{rad-rec}^{(6)} &= {\sum_a}\frac{m_e}{m_a} \bigg\{
	\left[\frac{3}{4} + \frac{6}{\pi^2}\zeta(3) - \frac{14}{\pi^2} - 2\ln 2\right] z_a^2 \pi \delta(\vb{r}_a)
	\nonumber\\&
	+ \left(\frac{2}{9}\pi^2 - \frac{70}{27}\right) \frac{z_a^2}{\pi} \delta(\vb{r}_a)
	\bigg\}.
	\label{eq:H_rad_r}
\end{align}

And the final set of contributions to $E^{(6)}$ consists of second-order perturbation corrections $E_\text{sec}$ due to the Breit--Pauli Hamiltonian $H^{(4)}$:
\begin{align}
	E_\text{sec}^{(6)} &= E_B + E_R + E_S, \label{eq:Esec} \\
	E_B &= \mel{\psi_0}{H_B Q (E_0 - H_0)^{-1} Q H_B}{\psi_0}, \label{eq:EB}\\
	E_R &= 2 \mel{\psi_0}{H_B Q (E_0 - H_0)^{-1} Q H_R}{\psi_0}, \label{eq:ER}\\
	E_S &= \mel{\psi_0}{H_S Q (E_0 - H_0)^{-1} Q H_S}{\psi_0}, \label{eq:ES-definition}
\end{align}
where
 $E_B$ is the contribution from the Breit interaction,
 $E_R$ is the relativistic correction due to retardation effects,
 $E_S$ is the contribution from spin-orbit interactions and
 $Q = 1 - \dyad{\psi_0}$ is the projection operator that excludes the reference state $\psi_0$.

The total $m\alpha^6$-order correction is obtained by summing the contributions from Eqs.~\eqref{eq:H6_rel}, \eqref{eq:H_H}, \eqref{eq:H_rad}, \eqref{eq:H_rad_r}, and \eqref{eq:Esec}:
\begin{align}\label{eq:E6}
	E^{(6)} = \expval{H_\text{rel}^{(6)} + H_H^{(6)} + H_\text{rad}^{(6)} + H_\text{rad-rec}^{(6)}} + E_\text{sec}^{(6)}.
\end{align}
Both the first-order perturbation contribution (the terms within angle brackets in Eq.~\eqref{eq:E6}) and the second-order contribution $E_\text{sec}^{(6)}$ exhibit native singularities due to the perturbation theory framework, which can and should be regularized. The details of this regularization are discussed in the next section.

\section{Elimination of Singularities and Regularization}

In quantum electrodynamics (QED), various regularization schemes are employed to handle singularities in theoretical calculations. As recently demonstrated by Korobov~\cite{Korobov2025reg,Korobov2025rec}, all these regularization methods yield unified finite expressions that are computationally tractable.

In our approach, we adopt the coordinate-space cutoff regularization scheme, where divergent integrals are rendered finite by introducing a minimal radial cutoff parameter~$r_0$ as the lower integration limit, see Eqs. \eqref{eq:integrals_r_0} and \eqref{eq:singular-integrals-cutoff}. The complete mathematical formulation and implementation details of this coordinate cutoff regularization procedure are thoroughly discussed in Refs.~\cite{Korobov2025reg,Korobov2025rec}. It is worthy to note that none of the distributions from Eq.~\eqref{eq:singular-integrals-cutoff} depends on the parameter $r_0$ and the numerical integration of these distributions based on prescriptions from \cite{Harris2004} is well defined and does not depend on the wave function expansion size nor particular choice of the Hylleraas basis or exponential basis.

The effective Hamiltonian that requires regularization is the relativistic correction \( H_{\text{rel}} \) of order \( m \alpha^6 \) in Eq.~\eqref{eq:H6_rel}. Additionally, the second-order corrections \( E_B \) and \( E_R \) given by Eqs.~\eqref{eq:EB} and \eqref{eq:ER}, respectively, must be addressed.

As can be seen from Appendix \ref{appdeix:effec-hamiltonian-modified}, singularities of the effective Hamiltonian can be isolated in the form of singular integrals of three types: $\langle V_a\rho_a\rangle$, $\langle V_a^3\rangle$ and $\langle \bm{\varepsilon}_a^2\rangle$. Then $\langle V_a\rho_a\rangle$ is transferred into the other two singular integrals using the identity \eqref{eq:va+rhoa} while $\langle V_a^3\rangle$ and $\langle \bm{\varepsilon}_a^2\rangle$ can be redefined as finite distributions using the coordinate cutoff regularization, see Eq.~(\ref{eq:singular-integrals-cutoff}).
In the following text, superscript $\langle\cdot\rangle^{(m)}$ denotes that the operator in the brackets is modified according to these prescriptions.
In the nonrecoil correction, these singularities disappear from the effective Hamiltonians, since they cancel each other out.
Thus, superscript $\langle\cdot\rangle'$ is used to indicate that regularized singular integrals of the above type are removed from the matrix element $\langle\cdot\rangle^{(m)}$.

In this section we consider only the two leading orders of the expansion in $m_e/m_a$, the higher order terms are ignored.

\subsection{Singularity in nonrecoil correction}

It has been demonstrated that the singularities in the $m\alpha^6$-order nonrecoil correction cancel out exactly in the hydrogen case~\cite{Korobov2007BO,Korobov2025reg}. For hydrogen molecular ions, the first-order perturbation contributions correspond to the effective Hamiltonians $\delta H_1$--$\delta H_3$ (see Eqs.~\eqref{eq:dH1}--\eqref{eq:dH3}). The singular part of the first-order contribution can be isolated as (see Eqs.(66)-(68) of Ref. \cite{Zhong2018NRQED}):
\begin{align}\label{eq:S123}
	S_{1,2,3} = \sum_a
	\bigg[&
	\frac{1}{32m_e^3}\left(1-\frac{5m_e}{m_a}\right)\langle \bm{\varepsilon}_a^2\rangle
	\nonumber\\&
	-\frac{1}{8m_e^2}\left(1-\frac{6m_e}{m_a}\right)\langle V_a^3\rangle
	\bigg]\,.
\end{align}

The nonrecoil second-order contribution is given by $E_B$ \eqref{eq:EB} and $E_S$ \eqref{eq:ES-definition}. To isolate the singularity in $E_B$, we introduce the transformation:
\begin{align}\label{eq:H_B_trans}
	H_B = H'_B + (E_0-H_0)U + U(E_0-H_0)\,,
\end{align}
where
\begin{align}\label{eq:U-definition}
	U = \lambda_1V_1 + \lambda_2V_2\,,\quad \lambda_a = -\frac{1}{4m_e}\left(1-\frac{3m_e}{m_a}\right)\,.
\end{align}

This allows $E_B$ to be expanded as:
\begin{align}\label{eq:EB-expand}
	E_B &= \mel{H'_B}{Q(E_0-H_0)Q}{H'_B} + \langle\{H_B,U\}\rangle
	\nonumber\\&
	 - 2\langle U\rangle \langle H_B\rangle - \langle U(E_0-H_0)U\rangle\,,
\end{align}
where the first term ($E'_B$) is numerically tractable, while the remaining three terms constitute a first-order contribution
\begin{align}\label{eq:HB-1st}
	H_B^\text{1st} = \{H_B,U\} - 2\langle{H_B}\rangle U - U(E_0-H_0)U\,.
\end{align}

The singular part of $H_B^\text{1st}$ can be extracted as $S_B$ :
\begin{align}\label{eq:SB}
	S_B= -\sum_a
	\bigg[&
	\frac{1}{32m_e^3}\left(1-\frac{5m_e}{m_a}\right)\langle \bm{\varepsilon}_a^2\rangle
	\nonumber\\&
	-\frac{1}{8m_e^2}\left(1-\frac{6m_e}{m_a}\right)\langle V_a^3\rangle
	\bigg]\,.
\end{align}
One can see that singularities in the nonrecoil correction after summation of all contributions are cancelled out $S_{1,2,3}+S_B\equiv0$.

Once singularities are removed, the modified effective Hamiltonian of the nonrecoil relativistic corrections is as
\begin{align}\label{eq:E6m_rel_nrec}
	{E'}_\text{rel}^{(6)} = \langle \delta H'_1+\delta H'_2+\delta H'_3+{H'}_B^\text{1st} \rangle\,.
\end{align}
%where the superscript '\emph{m}' stands for the modified effective Hamiltonian for the corresponding contributions.
The subtraction of the singularities and the determination of the finite expectation values for modified operators are moved to the Appendix; see sections \ref{app:divergent-matrix-elements} and \ref{appdeix:effec-hamiltonian-modified} for details.
Here we present the final results only:
\begin{align}\label{eq:H_1_reg}
	\langle\delta H'_1\rangle
	& =\frac{1}{16m_e^4}
	\bigg\{
	2E_0\langle p_e^4\rangle-\langle\{p_e^4,V_{12}\}\rangle
	-\sum_a
	\bigg[
	\langle\{p_e^4,V_a\}\rangle' \nonumber\\&
	+\frac{4m_e^2}{m_a}\left(E_0^2\langle p_a^2\rangle-2E_0\langle Vp_a^2\rangle+\langle Vp_a^2V\rangle'\right)
	\bigg]
	\bigg\}\,,
	\\
	\langle \delta H'_2\rangle
	& =\sum_a\bigg[-\frac{1}{32m_e^3}
	\bigg(
	\langle\bm{\varepsilon}_a\bm{\varepsilon}_b\rangle\mp\frac{m_e}{m_a}\langle\bm{\varepsilon}_a\bm{\varepsilon}_{12}\rangle
	\bigg)
	\nonumber\\&
	+\frac{3}{64m_e^4}\langle\{p_e^2,\rho_a\}\rangle'\bigg]\,, \label{eq:H_2_reg}
	\\
	\langle \delta H'_3\rangle
	&=\sum_a\frac{1}{8m_e^2m_a}(\mp)\langle\bm{\varepsilon}_a\bm{\varepsilon}_{12}\rangle\,, \label{eq:H_3_reg}
\end{align}
and
\begin{align}\label{eq:H_B_reg}
\begin{array}{@{}l}\displaystyle
	\langle {H'}_B^\text{1st}\rangle = -2\langle H_B\rangle\langle U\rangle
	+\frac{\lambda_1\lambda_2}{m_e}\langle\bm{\varepsilon}_1\bm{\varepsilon}_2\rangle
\\[2mm]\displaystyle%\hspace{22mm}
	-\sum_a\frac{\lambda_a}{8m_e^3}\left(\langle\{p_e^4,V_a\}\rangle'\!+\!2m_e\langle V_a\rho_b\rangle\!+\!2m_e\langle V_a\rho_a\rangle'\right)\,.
\end{array}
\end{align}
For convenience, we may not expand all the matrix elements into explicit formulas.

\subsection{Singularity in recoil correction}\label{sec:singularity_recoil}

We now proceed to isolate the singularities in the $m\alpha^6(m/M)$-order recoil correction, Eqs.~(\ref{eq:dH4})-(\ref{eq:dH6}) and (\ref{eq:H_H}). It should be noted that our previous consideration of this part~\cite{Zhong2018NRQED} was incomplete. Here we will use the results of Ref.~\cite{Korobov2025rec}, where it was shown that the singularities of recoil corrections from the low-energy region contribution are cancelled out when the singularity of the high-energy contribution $H_H^{(6)}$~\eqref{eq:H_H} is taken into account. The correct results are presented below; the details can be found in Appendix~\ref{appdeix:effec-hamiltonian-modified}.

Let us first consider $\delta H_4$, which can be explicitly expressed as
\begin{align} \label{eq:dH4_expansion_1st}
	\delta H_4 &= -\frac{1}{2m_e}\{E_0 - V, H_R\}
	\nonumber\\&
	= -\frac{E_0H_R}{m_e} - \sum_a \frac{z_ez_a}{2m_e^2m_a}\{V, p_e^i\mathcal{W}^{ij}_ap_a^j\}\,,
\end{align}
where the main part of the second term can be expanded as
\begin{align}
	\label{eq:dH4_vwp}
	&\langle \{V, p_e^i\mathcal{W}^{ij}_ap_a^j\} \rangle = 2\langle p_e^i\mathcal{W}^{ij}_aVp_a^j\rangle - \langle [p_e^i, V]\mathcal{W}^{ij}_ap_a^j\rangle  \nonumber\\&\qquad + \langle p_e^i\mathcal{W}^{ij}_a[p_a^j, V]\rangle \nonumber \\
	&= 2\langle p_e^i\mathcal{W}^{ij}_aVp_a^j\rangle
	- \langle \imath\varepsilon_a^i\mathcal{W}_a^{ij}p_a^j\rangle
	- \langle \imath\varepsilon_b^i\mathcal{W}_a^{ij}p_a^j\rangle_{b\neq a}
	\nonumber\\&\qquad
	- \langle p_e^i\mathcal{W}_a^{ij}\imath\varepsilon_a^j\rangle
	\mp \langle p_e^i\mathcal{W}_a^{ij}\imath\bm{\varepsilon}_{12}^j\rangle \nonumber\\
	&= 2\langle p_e^i\mathcal{W}^{ij}_aVp_a^j\rangle
	- \langle \imath\varepsilon_a^i\mathcal{W}_a^{ij}p_a^j\rangle
	- \langle p_e^i\mathcal{W}_a^{ij}\imath\varepsilon_a^j\rangle
	\,.
\end{align}
The matrix elements $\langle \imath\varepsilon_b^i\mathcal{W}_a^{ij}p_b^j\rangle{b\neq a}$ and $\langle p_e^i\mathcal{W}_a^{ij}\imath\varepsilon_{12}^j\rangle$ are numerically verified to be smaller than $10^{-20}$ and can therefore be removed from the final expressions safely.
The first term in the last line of Eq.~(\ref{eq:dH4_vwp}) is regular, while the other two are singular.
And the singularity in $\delta H_4$ can be written as
\begin{align}
	\label{eq:S4}
	S_4 = \sum_a\frac{1}{2m_e^2m_a}\left(\langle V_a\rho_a\rangle - \langle\bm{\varepsilon}_a^2\rangle\right)\,.
\end{align}
%The subscript $S$ on the matrix elements in Eq.~\eqref{eq:S4} denotes the singular part of the corresponding term.

For $\delta H_5$ one can directly obtain that the singular part may be written as
\begin{align}\label{eq:S5}
	S_5 = \sum_a\frac{1}{4m_e^2m_a}\langle \bm{\varepsilon}_a^2\rangle\,.
\end{align}

The derivation of the singular part in $\delta H_6$ is more complicated and we have moved it to the Appendix \ref{appdeix:effec-hamiltonian-modified}.
Here we present the final result only,
\begin{align}\label{eq:S6}
	S_6 = \sum_a\frac{1}{8m_e^2m_a}\left(\langle\bm{\varepsilon}_a^2\rangle-2m_e\langle V_a^3\rangle
	-\langle V_a\rho_a\rangle\right)\,.
\end{align}

Next we should consider the second-order contribution to the recoil correction: $E_R$~(\ref{eq:ER}). Applying the transformation (\ref{eq:H_B_trans}), $E_R$ takes the form
\begin{align}\label{eq:ER_expansion}
	E_R&=2\langle H'_BQ(E_0-H_0)^{-1}QH_R\rangle+\langle\{H_R,U\}\rangle-2\langle H_R\rangle \langle U\rangle\,,
\end{align}
where the first term ($E'_R$) is regular and the other two terms can be thought of as a modification of the effective Hamiltonian and considered as the first-order contribution
\begin{align}\label{eq:HR-1st}
	H_R^\text{1st}=\{H_R,U\}-2\langle H_R\rangle U\,.
\end{align}
The first term $H_R^\text{1st}$ is singular and can be considered in a similar way as $\delta H_4$.
Thus, the singular part of $E_R$ is
\begin{align}\label{eq:SR}
	S_R=\sum_a\frac{1}{4m_e^2m_a}\left(\langle\bm{\varepsilon}_a^2\rangle-\langle V_a\rho_a\rangle\right)\,.
\end{align}

Collecting all the singularities in Eqs. (\ref{eq:S4}), (\ref{eq:S5}), (\ref{eq:S6}) and (\ref{eq:SR}), one gets
\begin{align}\label{eq:S-rec-low}
	S_\text{rec}&=S_4+S_5+S_6+S_R
	\nonumber\\& 
	=\sum_a\frac{1}{8m_e^2m_a}\left(\langle\bm{\varepsilon}_a^2\rangle+\langle V_a\rho_a\rangle - 2m_e\langle V_a^3\rangle\right)\,.
\end{align}
Therefore, we should study integrals of those matrix elements $\langle\bm{\varepsilon}_a^2\rangle$, $\langle V_a^3\rangle$ and $\langle V_a\rho_a\rangle$, whose singular parts have been explicitly separated in Ref. \cite{Korobov2025reg} and Appendix \ref{app:cutoff}.

$S_\text{rec}$ will be regularized as ${(S_\text{rec})}_r$, namely regular part of $S_\text{rec}$,
\begin{align}\label{eq:S-rec-reg}
	{(S_\text{rec})}_r = \sum_a\frac{1}{8m_e^2m_a}\left(\langle\bm{\varepsilon}_a^2\rangle_r+\langle V_a\rho_a\rangle_r - 2m_e\langle V_a^3\rangle_r\right)\,.
\end{align}
by eliminating singularity of high-energy contribution $H_H^{(6)}$.
Subsequently, the regularization of recoil contribution will rise a term determined by the $\delta$-function distribution. In the cut-off regularization, it is expressed as
\begin{align}\label{eq:H-reg}
	H_r=\sum_aC_r\frac{z_a^3}{m_a}\pi\delta(\vb{r}_a)\,.
\end{align}
The coefficient $C_r=4\ln2-2$ is determined by comparing expectation value of regularized effective Hamiltonian with the {analytical} total $m\alpha^6$-oder correction in hydrogen atom, see Appendix \ref{app:H-ma6-rec}.

After regularization of all the contributions at $m\alpha^6(m/M)$ order, we obtain the effective Hamiltonians for the recoil relativistic corrections as follows:
\begin{align}\label{eq:E6m_rel_rec}
	{E'}^{(6)}_\text{rel-rec} &=  \langle \delta H'_4+\delta H'_5+\delta H'_6+H_r+ {H'}^\text{1st}_R\rangle
	\nonumber\\&\quad + {(S_\text{rec})}_r\,,
\end{align}
where
\begin{align}
	\langle \delta H'_4\rangle
	&=-\frac{E_0}{m_e}\langle H_R\rangle - \sum_a \frac{z_ez_a}{m_e^2m_a}
	\langle p_e^i\mathcal{W}^{ij}_aVp_a^j\rangle\,,\label{eq:H_4_reg}\\
	\langle \delta H'_5\rangle
	&= \sum_a\frac{1}{8m_e^2m_a}
	\bigg(
	\langle
	\vb{p}_eV_a^2\vb{p}_e
	\rangle
	+3\left\langle
	(\vb{p}_e\cdot\vb{r}_a)\frac{V_a^2}{r_a^2}(\vb{r}_a\cdot\vb{p}_e)
	\right\rangle
	\bigg)\,,\label{eq:H_5_reg}\\
	\langle \delta H'_6\rangle
	&=\sum_a\frac{1}{8m_e^2m_a}
	\bigg\{
	\left\langle \vb{p}_eV_a^2\vb{p}_e\right\rangle
	-3\left\langle (\vb{p}_e\cdot\vb{r}_a)\frac{V_a^2}{r_a^2}(\vb{r}_a\cdot\vb{p}_e)\right\rangle
	\nonumber\\&\quad
	-2z_ez_am_e\langle r_a\bm{\varepsilon}_a\bm{\varepsilon}_b\rangle
	\pm 8z_ez_am_e\bigg\langle
	[p_e^i,V]\mathcal{X}^{ij}_a\imath\bm{\varepsilon}_{12}^j
	\bigg\rangle
	\bigg\}\,,\label{eq:H_6_reg}
\end{align}
and
\begin{align}\label{eq:H_R_reg}
	\langle {H'}_R^\text{1st}\rangle
	=\sum_{a,b\neq a}
	-\frac{2z_ez_a}{m_e^2m_a}\langle p_e^i\mathcal{W}_a^{ij}Up_a^j\rangle
	-2\langle H_R\rangle \langle U\rangle
	\,.
\end{align}
It should be noted that the regular part ${(S_\text{rec})}_r$ of the divergent matrix elements appeared in $S_\text{rec}$ \eqref{eq:S-rec-low} should be retained after the regularization.

\subsection{the regularized effective Hamiltonian}

We now sum all contributions from Eqs.~\eqref{eq:H_H}, \eqref{eq:H_rad}, \eqref{eq:H_rad_r}, \eqref{eq:ES-definition}, \eqref{eq:E6m_rel_nrec}, \eqref{eq:H-reg} and \eqref{eq:E6m_rel_rec} to obtain the total corrections at orders $m\alpha^6$ and $m\alpha^6(m/M)$:
\begin{align}\label{eq:E6m}
	E^{(6)} &=
	{E'}_{\text{rel}}^{(6)} + {E'}_\text{sec}^{(6)} + E_{\text{rad}}^{(6)} + {E'}_{\text{rel-rec}}^{(6)} + E_{\text{rad-rec}}^{(6)}  \,.
%	+ E.
\end{align}
Comparing with previous $E^{(6)}$ of Eq. \eqref{eq:E6}, the relativistic correction due the effective Hamiltonian $H_\text{rel}^{(6)}$ has been regularized and split into nonrecoil correction $E_\text{rel}^{(6)}$ \eqref{eq:E6m_rel_nrec} and recoil correction $E_\text{rel-rec}^{(6)}$ \eqref{eq:E6m_rel_rec}.
It should be noted that the high-energy relativistic correction, so called pure recoil correction $H_H^{(6)}$, has been embraced in the contribution arising in the regularization $H_r$ \eqref{eq:H-reg}.
On the other hand,  the second-order corrections ${E}_\text{sec}^{(6)}$ is regularized as ${E'}_\text{sec}^{(6)}=E'_B+E_S+E'_R$ where $E'_B=\langle H'_BQ(E_0-H_0)^{-1}QH'_B\rangle$, $E'_R=2\langle H'_BQ(E_0-H_0)^{-1}QH_R\rangle$ and $E_S$ \eqref{eq:ES-definition} are regular enough for numerical study.

\section{Numerical Calculations and Conclusion}

\begin{table*}
	\centering
	\caption{Values of {regularized} divergent matrix elements for HD$^+$ rovibrational states (in atomic units).
		The numerical values given are exact to the last digit unless otherwise noted.
		Operators $\langle V_1\rho_1\rangle'$ are defined as in \eqref{eq:va+rhoa:reg}.}\label{tb:singular_operators}
	\begin{tabular}{ccccc}
	\hline\hline
	& (0,0) & (0,1) & (1,0) & (1,1) \\
	\hline
	$\langle V_1^3\rangle_r$&     1.6947~2934~9762(2) &  1.6601~8538~3679(7) & 1.6932~1792~262(3) & 1.6587~5603~832(2) \\
	$\langle V_2^3\rangle_r$&     1.6913~2284~4951 &  1.6567~3745~644(2)  & 1.6898~1215~596(7) & 1.6553~0853~833(3) \\
	$\langle \varepsilon_1^2\rangle_r$&    -1.4268~3109~35(2)   & --1.4012~5747~88(6) & --1.4256~45901(2) & --1.4001~32322(2) \\
	$\langle \varepsilon_2^2\rangle_r$&    -1.4254~5127~0134(7) & --1.3998~02588(1)   & --1.4242~65356(5) & --1.3986~76515(2) \\
	$\langle V_1\rho_1\rangle'$&    --2.9993~4511~0567 & --2.9150~8028~6650 & --2.996~5836~6165 & --2.9124~8207~422(6)  \\
	$\langle V_2\rho_2\rangle'$&    --2.9948~1780~0249 & --2.9104~8557~9425 & --2.992~0570~7310 & --2.9078~8765~6271(8) \\
	$\langle V_1^2\vb{p}_e^2\rangle'$&    --1.4813~9479~1135 & --1.4385~6298~1389 & --1.4804~2441~1796 & --1.4376~6199~5594\\
	$\langle V_2^2\vb{p}_e^2\rangle'$&    --1.4788~9959~4264 & --1.4360~3955~2156 & --1.4779~2998~6340 & --1.4351~3910~9269(5)\\
	$\langle V_1^2\vb{p}_1^2\rangle'$&     5.9701~0577~1673  & 20.3159~9546~6009 & 6.6687~9581~7755(3) & 20.9624~8160~67(2)\\
	$\langle V_2^2\vb{p}_2^2\rangle'$&     5.9623~8727~4107  & 20.2891~4197~9807 & 6.6609~00330(6)     & 20.9354~8110(1) \\
	$\langle V\vb{p}_1^2V\rangle'$&    12.7118~9202~2087 & 39.2947~8595~0492 & 14.0043~4653~5632(3) & 40.4898~3443~34(2) \\
	$\langle V\vb{p}_2^2V\rangle'$&    12.7102~4302~4887 & 39.2859~6578~8716 & 14.0030~73800(6)     & 40.4814~0269(1)    \\
	$\langle \{\vb{p}_e^2,\rho_1\}\rangle'$&    18.2471~9863~2286 & 17.7214~5445~3675 & 18.2305~9594~1969(2) & 17.7058~8161~45(4) \\
	$\langle \{\vb{p}_e^2,\rho_2\}\rangle'$&    18.2147~1435~2833 & 17.6887~0915~2486 & 18.1981~2087~512(5)  & 17.6731~4265~18(2) \\
	$\langle \{\vb{p}_e^4,V_1\}\rangle'$&     8.0683~0063~2230 & 7.8427~3701~1450 & 8.0647~8568~0309(5) & 7.8396~3767~03(2) \\
	$\langle \{\vb{p}_e^4,V_2\}\rangle'$&     8.0437~1320~4651 & 7.8181~1053~5982 & 8.0402~1158~890(2)  & 7.8150~2238~49(1) \\
	\hline\hline
	\end{tabular}
\end{table*}

\begin{table*}
	\centering
	\caption{Values of regular matrix elements for HD$^+$ rovibrational states (in atomic units).
	The numerical values given are exact to the last digit unless otherwise noted.}\label{tb:regular_operators}
    \hspace*{-10mm}
	\begin{tabular}{ccccc}
		\hline\hline
		& (0,0) & (0,1) & (1,0) & (1,1) \\
		\hline
$E_0$&    --0.5978~9796~8610 & --0.5891~8182~9561 & --0.5976~9812~8194 & --0.5889~9111~1996 \\
$\langle U\rangle$&     0.4214~4656~2271 & 0.4129~4440~6667 & 0.4212~2454~0596 & 0.4127~3112~3370 \\
$\langle H_B\rangle$&   --0.1366~0153~9235 & --0.1338~7800~1186 & --0.1364~6376~9526 & --0.1337~4881~366(2) \\
$\langle H_R\rangle$&   --0.0004~7880~1585 & --0.0004~6806~9095 & --0.0004~7839~1583 & --0.0004~6768~4165 \\
$\langle V_1\rho_2\rangle$&     1.2932~5736~5489 & 1.2445~4449~4788 & 1.2907~9418~6712(2) & 1.2421~6853~875(1) \\
$\langle V_2\rho_1\rangle$&     1.2949~7113~2033 & 1.2462~3915~8675 & 1.2925~0636~9768    & 1.2438~6174~637(5) \\
$\langle \vb{p}_1^2\rangle$&     5.9805~4596~1465 & 16.0975~3333~1318 & 6.4633~5834~8141  & 16.5472~3819~2988 \\
$\langle \vb{p}_2^2\rangle$&     5.9782~9548~7288 & 16.0925~0308~7347 & 6.4609~7631~9844  & 16.5420~8531~8687 \\
$\langle V\vb{p}_1^2\rangle$&    --8.3947~4694~8040 & --20.3714~7544~2789 & --8.9728~6271~9712   & --20.9059~8595~098(5) \\
$\langle V\vb{p}_2^2\rangle$&    --8.3901~0978~4358 & --20.3635~0767~2653 & --8.9680~6789~950(3) & --20.8978~7211~8122(5) \\
$\langle \vb{p}_e^4\rangle$&     6.3001~9994~7706 & 6.1590~2235~2359 & 6.2944~5074~6088(5) & 6.1535~9974~587(8) \\
$\langle \{\vb{p}_e^4,V_{12}\}\rangle$&     6.2640~6896~7646 & 6.0337~6087~260(1) & 6.2520~0879~745(2) & 6.0221~1748~900(3) \\
$\langle \bm{\varepsilon}_1\cdot\bm{\varepsilon}_2\rangle$&    --0.0544~9531~3576 & --0.0508~0894~3618 & --0.0543~7418~1182 & --0.0506~9770~3844 \\
$\langle \bm{\varepsilon}_1\cdot\bm{\varepsilon}_{12}\rangle$&    --0.0615~4079~8892 & --0.0590~6505~7552 & --0.0613~4476~1159 & --0.0588~7417~7983 \\
$\langle \bm{\varepsilon}_2\cdot\bm{\varepsilon}_{12}\rangle$&     0.0615~4700~6305 &  0.0590~7129~4583 & 0.0613~5094~1177 & 0.0588~8038~6718 \\
$\langle \vb{p}_eV_1^2\vb{p}_e\rangle$&     1.5179~5031~9432 & 1.4765~1730~5261 & 1.5161~5924~9854 & 1.4748~2007~862(6) \\
$\langle \vb{p}_eV_2^2\vb{p}_e\rangle$&     1.5159~1820~5985 & 1.4744~4602~7269 & 1.5141~2708~6759(5) & 1.4727~4854~7002(6) \\
$\langle (\vb{p}_e\cdot\vb{r}_1)(V_1^2/r_1^2)(\vb{r}_1\cdot\vb{p}_e)\rangle$&     1.3200~5514~4383 & 1.2895~5695~8897 & 1.3187~0656~3776 & 1.2882~8015~190(8) \\
$\langle (\vb{p}_e\cdot\vb{r}_2)(V_2^2/r_2^2)(\vb{r}_2\cdot\vb{p}_e)\rangle$&     1.3178~7214~0481 & 1.2873~3731~9221 & 1.3165~2379~6039(5) & 1.2860~6053~815(1) \\
$\langle z_ez_1p_e^i\mathcal{W}_1^{ij}Vp_1^j\rangle$&    --1.2281~0539~4297 & --1.2008~2524~9325 & --1.2270~8820~5793 & --1.1998~6987~832(5) \\
$\langle z_ez_2p_e^i\mathcal{W}_2^{ij}Vp_2^j\rangle$&    --1.2241~3289~5621 & --1.1936~4318~5135 & --1.2229~7915~8165(5) & --1.1925~6219~346(2) \\
$\langle z_ez_1p_e^i\mathcal{W}_1^{ij}Up_1^j\rangle$&     0.3795~4851~4771  & 0.3700~4007~8771 & 0.3791~6464~1365 & 0.3696~7653~009(1) \\
$\langle z_ez_2p_e^i\mathcal{W}_2^{ij}Up_2^j\rangle$&     0.3784~3817~9175  & 0.3682~5701~6903 & 0.3780~2354~8783(1) & 0.3678~6502~5691(6) \\
$\langle z_ez_1r_1\bm{\varepsilon}_1\cdot\bm{\varepsilon}_2\rangle$&     0.0169~4925~9569 & 0.0148~8857~1832 & 0.0169~8530~2788 & 0.01492~6808~1686 \\
$\langle z_ez_1r_2\bm{\varepsilon}_1\cdot\bm{\varepsilon}_2\rangle$&     0.0169~1867~6737 & 0.0148~5875~0542 & 0.0169~5477~8808 & 0.01489~7043~4046 \\
$\langle z_ez_1[p_e^i,V]\mathcal{X}_1^{ij}\imath\varepsilon_{12}^j\rangle$&     0.0516~8603~2319 & 0.0495~1598~2047 & 0.0515~5049~9805 & 0.0493~8474~3758 \\
$\langle z_ez_2[p_e^i,V]\mathcal{X}_2^{ij}\imath\varepsilon_{12}^j\rangle$&   --0.0517~2674~9355 & --0.0495~5592~2695 & --0.0515~9115~4163 & --0.0494~2462~4074 \\
		\hline\hline
	\end{tabular}
\end{table*}

\begin{table*}
	\centering
	\caption{Expectation values of modified effective Hamiltonians for HD$^+$  rovibrational states  (in atomic units).
	The numerical values given are exact to the last digit unless otherwise noted.}\label{tb:eff_Ham}
	\begin{tabular}{cccccc}
	\hline\hline
   & Eqs. & (0,0) & (0,1) & (1,0) & (1,1)\\
	\hline
	$\langle\delta H'_1\rangle$& \eqref{eq:H_1_reg}&   --1.8703~480(6) & --1.8137~755(5) & --1.8687~273(6) &  --1.8122~633(5)\\
	$\langle\delta H'_2\rangle$&\eqref{eq:H_2_reg}&     1.7125~566(5) &   1.6630~255(5) &   1.7109~929(5) &    1.6615~589(5)\\
	$\langle\delta H'_3\rangle$&\eqref{eq:H_3_reg}&   --0.0000~0628~5744(2) & --0.0000~0603~2892(2) & --0.0000~0626~5720(2) & --0.0000~0601~3395(2)\\
	$\langle\delta {H'}_B^\text{1st}\rangle$&\eqref{eq:H_B_reg}&     0.4020~166(1) & 0.3880~342(1) & 0.4016~656(1) & 0.3877~119(1)\\
	${E'}_\text{rel}^{(6)}$&\eqref{eq:E6m_rel_nrec}&     0.2442~189(8) & 0.2372~782(7) & 0.2439~249(8) & 0.2370~001(7)\\
	&&&&&\\
	$\langle\delta H'_4\rangle$&\eqref{eq:H_4_reg}&     0.0007~150(4) & 0.0007~015(4) & 0.0007~144(4) & 0.0007~010(4) \\
	$\langle\delta H'_5\rangle$&\eqref{eq:H_5_reg}&     0.0005~5890~9980  & 0.0005~4532~3594  & 0.0005~5831~3896 & 0.0005~4475~9030\\
	$\langle\delta H'_6\rangle$&\eqref{eq:H_6_reg}&   --0.0002~1032~7438  & --0.0002~0656~2370& --0.0002~1021~536(2) & --0.0002~0645~96(2)\\
	$\langle {H'}_R^\text{1st}\rangle$&\eqref{eq:H_R_reg}&  --0.0002~1512~4135  & --0.0002~1550~1367(2) & --0.0002~1500~833(1) & --0.0002~1538~1914(5)\\
	${(S_\text{rec})}_r$&\eqref{eq:S-rec-reg}&     --0.0010~2281~4620 & --0.0010~0094~4942 & --0.0010~2191~0494 & --0.0010~0008~9977\\
	$\langle H_r\rangle$&\eqref{eq:H-reg}&     0.0004~1079~5750 & 0.0004~0137~2924 & 0.0004~1042~9068 & 0.0004~0102~6584\\
	${E'}_\text{rel-rec}^{(6)}$&\eqref{eq:E6m_rel_rec}&     0.0002~364(4)    & 0.0002~251(4) & 0.0002~360(4) & 0.0002~248(4)\\
	\hline\hline
	\end{tabular}
\end{table*}

\begin{table*}
	\centering
	\caption{QED corrections of order $m\alpha^6$ for HD$^+$ rovibrational transitions $(L,v)\rightarrow(L',v')$, in kHz.}
		\label{tb:HD+-(00-01)-contibutions}
	\begin{tabular}{lddd}
	\hline\hline
     & \multicolumn{1}{c}{$(0,0)\rightarrow(0,1)$} & \multicolumn{1}{c}{$(0,0)\rightarrow(1,0)$} & \multicolumn{1}{c}{$(0,0)\rightarrow(1,1)$} \\
%     & & & \\
	\hline
	$\Delta E_\text{rad}^{(6)}$ & -1735.4057 & -67.5480 &-1799.2044\\
	$\Delta {E'}_\text{rel}^{(6)}$ & -129.500(20) & -5.485(21) & -134.689(20)\\
	$\Delta (E'_B+E'_S)$ & 153.998(26) & 7.353(26) & 160.864(26)\\
	$\Delta E_\text{rad-rec}^{(6)}$ &  0.3105 & 0.0121 & 0.3219\\
	$\Delta {E'}_\text{rel-rec}^{(6)}$ & -0.211(11) & -0.007(11) & -0.216(11)\\
	$\Delta E'_R$ & 0.5868(3) & 0.0265(3) & 0.6112(3)\\
	\hline
	$\Delta E_\text{total}$ & -1710.{221}(35) & -65.{649}(35) & -177{2}.{313}(35) \\
	~~\cite{Korobov2017ppt} & -1708.9(1) & & \\
	~~\cite{Kortunov2021}& & & -1770.95(10)\footnote{The uncertainty keeps the value estimated in Ref. \cite{Korobov2017ppt}.}  \\
	\hline\hline
	\end{tabular}
\end{table*}

The matrix elements for the operators that appear in the $m\alpha^6$-order correction $E^{(6m)}$, see Eq. \eqref{eq:E6m}, are calculated using the Hylleraas variational expansion of the wave function:
\begin{align}\label{eq:basis-function-hylleraas-1}
	\phi_{ijk}^{l_1l_2L}=r_1^{i}r_2^{j}r_{12}^{k}e^{-\alpha{r}_1-\beta{r}_2}Y^{l_1l_2}_{LM}(\hat{r}_1,\hat{r}_2)\,.
\end{align}
In the above, $i,j,k$ are integers satisfying $i+j+k\leq\Omega$ with $\Omega$ being a parameter defining the size of the basis set, $\alpha$ and $\beta$ are nonlinear parameters, and $Y^{l_1l_2}_{LM}(\hat{r}_1,\hat{r}_2)$ is the vector coupled production of spherical harmonics.
The nonrelativistic wave functions and {corresponding} energies are calculated in the center-of-mass frame.
More details on the construction of the basis set for HD$^+$ may be found in Refs.~\cite{Yan2003,Yan2007,Zhong2015CPB}.
Most integrals can be calculated analytically using Perkins' expansion for $r_{12}^k$ \cite{Yan1996}.
The procedure for handling singular integrals that appear in the evaluation can be found in Ref.~\cite{Yan1994,Qi2023SI}.
The divergent matrix elements $\langle V_a^3\rangle$, $\langle\bm{\varepsilon}_a^2\rangle$ and $\langle V_a\rho_a\rangle$ are calculated in the regularized form as defined in Eq. \eqref{eq:singular-integrals-cutoff}.

Table~\ref{tb:singular_operators} shows the values of regularized divergent matrix elements that appear in the modified effective Hamiltonian, while Table~\ref{tb:regular_operators} contains the numerical values for the finite matrix elements in Eq.~\eqref{eq:E6m}, including the nonrelativistic energy $E_0$, the potential $U$, Eq.~\eqref{eq:U-definition}, the Breit Hamiltonians $H_B$ and $H_R$, Eqs.~\eqref{eq:ma4-HB} and \eqref{eq:ma4-HR}.
The uncertainties presented in these tables arise mainly from limitations of the basis set, which are denoted as numerical uncertainties.

Table~\ref{tb:eff_Ham} presents the expectation values of the modified effective Hamiltonians that appear in the nonrecoil and recoil relativistic corrections ${E'}_\text{rel}^{(6)}$ and ${E'}_\text{rel-rec}^{(6)}$.
Because of the regularization and transformation applied to the relativistic corrections, these modified effective Hamiltonians are derived by neglecting recoil corrections beyond the leading orders.
Consequently, our theoretical framework exhibits a relative uncertainty of $\mathcal{O}((m_e/m_p)^2)$ for effective Hamiltonians $\delta H'_1$, $\delta H'_2$, $\delta H'_3$, $\delta H'_4$ and ${H'}_B^\text{1st}$, even though our numerical precision is significantly higher.
Therefore, the uncertainties in the $m\alpha^6$ order nonrecoil correction ${E'}_\text{rel}^{(6)}$ and recoil correction ${E'}_\text{rel-rec}^{(6)}$ arise mainly from the {theoretical} uncertainty caused by the yet uncalculated higher-order recoil corrections.

The regularized second-order contributions $E'_B$, $E'_R$ and $E_S$ have recently been computed in~\cite{Korobov2025sec} for the rovibrational states of H$_2^+$ and HD$^+$.
These second-order terms were evaluated using the algebraic technique developed by Dalgarno and Lewis~\cite{Dalgarno1955}. The numerical computation of these terms remains particularly challenging. As reported in~\cite{Korobov2025sec}, the relative numerical precision of these second-order contributions is estimated to be approximately $10^{-5}$.
In this work, we use the values of $E'_B$, $E'_R$ and $E_S$ from Ref.~\cite{Korobov2025sec} for the second-order contributions to calculate the rovibrational transition frequencies.

The $m\alpha^6$ order corrections to transition frequencies in the HD$^+$ ion are presented in Table~\ref{tb:HD+-(00-01)-contibutions}. The uncertainty in the nonrecoil relativistic correction $\Delta {E'}_\text{rel}^{(6)}$ and the recoil relativistic correction $\Delta {E'}_\text{rel-rec}^{(6)}$ are estimated to be 20~Hz and 11~Hz respectively, primarily due to neglect of the higher-order recoil effects. The uncertainty due to the second-order contribution $\Delta (E'_B+E_S)$ is about 26~Hz, which is estimated by the last significant digit of the corresponding values of $E'_B$ and $E_S$. Therefore, a total uncertainty of 35~Hz in the transitions arises mainly from the uncertainty in the second-order corrections. The resulting $m\alpha^6$-order correction to the $(0,0)\!\to\!(0,1)$ transition is determined to be $-1710.{221}(35)~\text{kHz}$, the uncertainty of which is three orders of magnitude smaller than the value reported in previous theoretical studies \cite{Korobov2017ppt,Korobov2021}. 
As is seen, there is a discrepancy of approximately {1.4}~kHz, which is {about 40} times larger than the present total uncertainty, between the obtained values and the previous theory on transitions $(0,0)\!\to\!(0,1)$ and $(0,0)\!\to\!(1,1)$ as shown in Table~\ref{tb:HD+-(00-01)-contibutions}.
The discrepancy between adiabatic approximation and the nonrelativistic three-body solution appears already in the second-order corrections. For example,  the three-body calculation of $E_B$ yields $-0.3009345$ on the ground state of HD$^+$, while the adiabatic approximation gives $-0.302712$ .
We attribute this to the limitations of the adiabatic BO approximation.

\begin{table}\centering
	\caption{Comparison of $m\alpha^6$ order contributions between this work and previous BO based results on HD$^+$ fundamental transition $(0,0)\rightarrow(0,1)$, in kHz.}
	\label{tb:comparison-ma6}
	\begin{tabular}{ldd}
		\hline
		 & \multicolumn{1}{c}{this work} & \cite{Korobov2017ppt} \\
		\hline
		$\Delta E^{(6)}_\text{rad}$ & -1735.4057 & -1735.4 \\
		$\Delta {E}^{(6)}_\text{rel-nrec}$ & 24.498(33)  & {26}.{0} \\
		$\Delta E^{(6)}_\text{recoil}$ & 0.100(11)  & 0.{49(2)} \\
		$\Delta E'_R$ & 0.5868(3) & - \\
		\hline
		$\Delta E_\text{total}$ & -1710.221(35) & -1708.9(1) \\
		\hline
	\end{tabular}
\end{table}

{We also compare the $m\alpha^6$ order contributions from this work with those based on the BO approximation, as shown in Table \ref{tb:comparison-ma6}. 
The total nonrecoil relativistic correction $\Delta E^{(6)}_\text{rel-nrec}$ consists of a first-order correction $\Delta {E'}^{(6)}_\text{rel}$ and a second-order correction $\Delta (E'_B+E'_S)$.
The recoil correction $\Delta E^{(6)}_\text{recoil}$ includes the recoil radiative contribution $\Delta E^{(6)}_\text{rad-rec}$ and the recoil relativistic correction $\Delta E^{(6)}_\text{rel-rec}$.
One can observe the differences between $\Delta E^{(6)}_\text{rel-nrec}$ and $\Delta E^{(6)}_\text{recoil}$, as well as the absence of the second-order contribution $\Delta E'_R$ in BO theory.}

In conclusion, we present numerical calculations for the first-order corrections at $m\alpha^6$- and $m\alpha^6(m/M)$-order in the three-body formalism for the hydrogen molecular ions. It relies heavily on the coordinate cutoff regularization. The regularization procedure is presented in detail, especially for the recoil corrections. Matrix elements appearing in the regularized effective Hamiltonians are calculated to high precision based on the Hylleraas basis set. The expectation values of effective Hamiltonians are presented with a relative uncertainty of $\mathcal{O}((m_e/m_{p})^2)$. In combination with the numerical results for the second-order corrections~\cite{Korobov2025sec}, the total contribution of the $m\alpha^6$-order to the fundamental transitions in HD$^+$ is obtained with precision of $35~\text{Hz}$, which is three orders of magnitude better than in the previous work.
As stated in \cite{Korobov2017ppt}, the largest contribution to the uncertainty comes from the $m\alpha^8$ order correction, which is four times larger than the uncertainty in the $m\alpha^6$ order correction.
Nevertheless, the results obtained in this work allow us to state that in theoretical studies of the spectra of vibrational transitions in molecular hydrogen ions, a relative accuracy of $10^{-12}$ or higher can be achieved.

\begin{acknowledgments}
This research was supported by the National Natural Science Foundation of China (NSFC) under Grants Nos. 12393821.
ZXZ acknowledges support from the Innovation Fund for Scientific and Technological Personnel of Hainna Province (Grant No. 225RC636).
Chun Li was supported in part by the National Natural Science Foundation of China Grant (No. 11871271).
T. Y. Shi was supported by the Strategic Priority Research Program of the Chinese Academy of Sciences (Grant Nos.XDB0920100 and  XDB0920101), the National Natural Science Foundation of China (Grant No.12274423).
Authors would like to thank helpful comments from Z.-C. Yan of UNB and J.-Ph. Karr of LKB.
\end{acknowledgments}

\appendix

\section{Coordinate cutoff regularization}\label{app:cutoff}

In the coordinate cutoff regularization, the divergent integrals can be integrated by setting a lower limit for integration over the radial variable to some small parameter $r_0$. The finite distribution for $1/r^3$ integral was introduced by Araki \cite{Araki1957} and Sucher \cite{Sucher1958} in connection with the $ m\alpha^5$ order correction for the helium atom. The general analysis of the cutoff regularization for the $m\alpha^6$ order relativistic {corrections} has been given in Ref.~\cite{Korobov2025reg}, Section~IV.A.

Here we present a brief summary of the main features of the cutoff regularization for bound states. In our case, all singularities are separated out in the form of singular integrals: $ \langle V^3 \rangle $, $ \langle \bm{\varepsilon}^2 \rangle $, $\langle V\imath\bm{\varepsilon}\vb{p}_e\rangle$ and $ \langle V\rho \rangle $.
The divergent part of those integrals may be expressed as follows:
\begin{align}\label{eq:integrals_r_0}
\begin{array}{@{}l}\displaystyle
	\langle V^3\rangle_{r_0} = \left\langle\frac{(z_ez_N)^3}{r^3}\right\rangle_{r_0}
\\[3mm]\displaystyle\hspace{11mm}
	=(\ln r_0+\gamma_E)(z_ez_N)^2\langle 4\pi\delta(\vb{r})\rangle+\text{const.}\,,
\\[3mm]\displaystyle
	\langle\bm{\varepsilon}^2\rangle_{r_0} = \left\langle\frac{(z_ez_N)^2}{r^4}\right\rangle_{r_0}
\\[3mm]\displaystyle\hspace{9mm}
	=\frac{(z_ez_N)^2}{r_0}\langle4\pi\delta(\vb{r})\rangle
	+(\ln r_0\!+\!\gamma_E)(z_ez_N)^2\langle4\pi\delta'(\vb{r})\rangle+\text{const.}\,,
\\[3mm]\displaystyle
	\langle V\imath\bm{\varepsilon}\vb{p}\rangle_{r_0} = \frac{\ln r_0\!+\!\gamma_E}{2}(z_ez_N)^2\langle4\pi\delta'(\vb{r})\rangle+\text{const.}\,,
\\[3mm]\displaystyle
	\langle V\rho\rangle_{r_0}\equiv\langle\bm{\varepsilon}^2\rangle_{r_0}
	-2\langle V\imath\bm{\varepsilon}\vb{p}\rangle_{r_0}
\\[3mm]\displaystyle\hspace{11mm}
	=\frac{(z_ez_N)^2}{r_0}\langle4\pi\delta(\vb{r})\rangle-(z_ez_N)^2\langle4\pi\delta'(\vb{r})\rangle\,,
\end{array}
\end{align}
where $\gamma_E$ is the Euler gamma constant and
\begin{align*}
	\mel{\phi_1}{\delta'(\bm{r})}{\phi_2} &= \mel{\phi_1}{\frac{\bm{r}}{r} \nabla \delta(\bm{r})}{\phi_2}
	\nonumber\\&
	= -\mel{\partial_r \phi_1}{\delta(\bm{r})}{\phi_2} - \mel{\phi_1}{\delta(\bm{r})}{\partial_r \phi_2}\,.
\end{align*}
In the above formulas, $ V = z_e z_N / r $ is the Coulomb interaction between the electron and the nucleus, where $ z_e $ and $ z_N $ are the charges of the electron and the nucleus, respectively.

In the coordinate cutoff regularization, the finite integrals for these matrix elements are defined as follows:
\begin{align}\label{eq:singular-integrals-cutoff}
\begin{array}{@{}l}\displaystyle
	\langle V \rangle_r = (z_e z_N)^3 \lim_{r_0 \rightarrow 0} \left\{ \left\langle \frac{1}{r} \right\rangle_{r_0} + (\ln r_0 + \gamma_E) \langle 4\pi \delta(\bm{r}) \rangle \right\},
\\[3mm]\displaystyle
	\langle \bm{\varepsilon}^2 \rangle_r = (z_e z_N)^2 \lim_{r_0 \rightarrow 0} \Bigg\{ \left\langle \frac{1}{r} \right\rangle_{r_0} - \bigg[ \frac{\langle 4\pi \delta(\bm{r}) \rangle}{r_0}
\\[3mm]\displaystyle\hspace{35mm}
	+ (\ln r_0 + \gamma_E) \langle 4\pi\delta'(\bm{r}) \rangle \bigg] \Bigg\},
%	\langle V\imath\bm{\varepsilon}\vb{p}\rangle_r & = \lim_{r_0 \rightarrow 0}
%	\left\{
%		\langle V\imath\bm{\varepsilon}\vb{p}\rangle_{r_0}
%		-(z_ez_N)^2\frac{\ln r_0+\gamma_E}{2}\langle4\pi\delta'(\vb{r})\rangle
%	\right\}\,,\\
\\[3mm]\displaystyle
	\langle V\rho\rangle_r %= \langle\bm{\varepsilon}^2\rangle_r-2\langle V\imath\bm{\varepsilon}\vb{p}\rangle_r
	=-(z_ez_N)^2\langle4\pi\delta'(\vb{r})\rangle\,.
\end{array}
\end{align}

The numerical evaluation of the divergent integrals in Eq.~(\ref{eq:singular-integrals-cutoff}) for the Hylleraas variational basis functions used in our work is discussed in Ref.~\cite{Harris2004}.

\section{Divergent matrix elements}\label{app:divergent-matrix-elements}

In order to express $H^{(6)}_{\rm rel}$ in Eq.~(\ref{eq:H6_rel}) in terms of finite matrix elements, we transform the effective Hamiltonian and the second-order corrections into a set of regular and divergent (but redefined in the finite form) matrix elements, such as $\langle V_a \vb{p}_e^2 V_a \rangle$, $\{\vb{p}_e^4, V_a\}$, and $\langle \vb{p}_e^2 V_a \vb{p}_e^2 \rangle$, etc.
Many of these matrix elements were already considered in Ref.~\cite{Zhong2018NRQED}; here we write out {explicitly} our previously obtained results and derive additional new expressions for matrix elements that appear in this work.

\subsection{Matrix elements $\langle V_a^2\vb{p}_e^2\rangle$, $\langle V_a^2\vb{p}_a^2\rangle$, $\langle V_a\vb{p}_e^2V_a\rangle$, $\langle V_a\vb{p}_a^2V_a\rangle$ and $\langle{V}\vb{p}_a^2V\rangle$}

In this section we will frequently use the following expression for $\vb{p}_e^2\psi_0$:
\begin{align}\label{eq:SE_trans}
	\vb{p}_e^2\psi_0 = \left[2m_e(E_0-V)-\sum_a\frac{m_e}{m_a}\vb{p}_a^2\right]\psi_0\,.
\end{align}
which is equivalent to the Schr\"{o}dinger equation (\ref{eq:H0}).

The finite (regularized) distribution for the matrix element $\langle V_a^2\vb{p}_e^2\rangle$ can be determined as follows
\begin{comment}
\begin{align}\label{eq:vas+pes}
\begin{array}{@{}l}\displaystyle
		\langle V_a^2\vb{p}_e^2\rangle^{(m)} =
        -2\mu_a\langle V_a^3\rangle_r
		+\mu_a \bigg\{2\left[E_0\langle V_a^2\rangle-\langle V_a^2(V_b+V_{12})\rangle\right]
\\[3mm]\displaystyle\hspace{45mm}
		-\frac{1}{m_a}\langle V_a^2(\vb{p}_a^2-\vb{p}_e^2)\rangle
		-\frac{1}{m_b}\langle V_a^2\vb{p}_b^2\rangle
		\bigg\}\,,
\end{array}
\end{align}
\end{comment}
\begin{align}\label{eq:vas+pes}
	\langle V_a^2\vb{p}_e^2\rangle^{(m)} = -2\mu_a\langle V_a^3\rangle_r + \langle V_a^2\vb{p}_e^2\rangle'\,,
\end{align}
where the first term on the right-hand side is regularized divergent, and the second term is finite
\begin{align}\label{eq:vas+pes:reg}
	\langle V_a^2\vb{p}_e^2\rangle' &= \mu_a \bigg\{2\left[E_0\langle V_a^2\rangle-\langle V_a^2(V_b+V_{12})\rangle\right]
	\nonumber\\&\qquad
	-\frac{1}{m_a}\langle V_a^2(\vb{p}_a^2-\vb{p}_e^2)\rangle
	-\frac{1}{m_b}\langle V_a^2\vb{p}_b^2\rangle
	\bigg\}\,.
\end{align}
In $\langle V_a^2\vb{p}_e^2\rangle'$,  indexes $a$ and $b$ stand for the two nuclei with $b\neq{a}$, and $\mu_a=m_am_e/(m_a+m_e)$ are the reduced masses.

The finite matrix element for $\langle V_a^2\vb{p}_a^2\rangle$ then can be determined
\begin{equation}\label{eq:vas+pas}
	\langle V_a^2\vb{p}_a^2\rangle^{(m)} = -2\mu_a\langle V_a^3\rangle_r + \langle V_a^2\vb{p}_a^2\rangle'\,,
\end{equation}
where
\begin{equation}\label{eq:vas+pas:reg}
	\langle V_a^2\vb{p}_a^2\rangle' = \langle V_a^2\vb{p}_e^2\rangle' + \langle V_a^2(\vb{p}_a^2-\vb{p}_e^2)\rangle\,.
\end{equation}

The matrix element $\langle V_a\vb{p}_e^2V_a\rangle$ may be expressed as follows (see Eq. (A6) in Ref. \cite{Zhong2018NRQED})
\begin{align}\label{eq:va+pes+va}
	\langle V_a\vb{p}_e^2V_a\rangle^{(m)} = \langle\bm{\varepsilon}_a^2\rangle_r-2\mu_a\langle V_a^3\rangle_r + \langle V_a^2\vb{p}_e^2\rangle'\,.
\end{align}
The matrix element $\langle V_a\vb{p}_a^2V_a\rangle$ is expressed in the similar form
\begin{align}
	\langle V_a\vb{p}_a^2V_a\rangle^{(m)} =\langle\bm{\varepsilon}_a^2\rangle_r-2\mu_a\langle V_a^3\rangle_r+\langle V_a^2\vb{p}_a^2\rangle'\,.
\end{align}
And the cross-term
\begin{align}\label{eq:va+pes+vb}
	\langle V_a\vb{p}_e^2V_b\rangle = \langle \bm{\varepsilon}_a\bm{\varepsilon}_b\rangle + \langle V_aV_b\vb{p}_e^2\rangle\,.
\end{align}
In the latter case all the operators are finite.

Now we may get the matrix element $\langle V\vb{p}_a^2V\rangle$:
\begin{align}\label{eq:v+pas+v}
	\langle V\vb{p}_a^2V\rangle^{(m)} = \langle\bm{\varepsilon}_{12}^2\rangle_r+\langle\bm{\varepsilon}_a^2\rangle_r-2\mu_a\langle V_a^3\rangle_r
	+ \langle V\vb{p}_a^2V\rangle'\,,
\end{align}
where
\begin{align}\label{eq:v+pas+v:reg}
	\langle V\vb{p}_a^2V\rangle' = \langle{V}_a^2\vb{p}_a^2\rangle'\pm2\langle\bm{\varepsilon}_a\bm{\varepsilon}_{12}\rangle+\langle(V^2-V_a^2)\vb{p}_a^2\rangle\,.
\end{align}
In the above derivation, singular integrals involving $r_{12}$ appear implicitly in the form $\langle\bm{\varepsilon}_{12}^2\rangle+\langle V_{12}^2\vb{p}_a^2\rangle$, which can be transferred as (see Eq. (21) of Ref. \cite{Korobov2025reg})
\begin{align*}
	\langle\bm{\varepsilon}_{12}^2\rangle+\langle V_{12}^2\vb{p}_a^2\rangle=\langle V_{12}\rho_{12}\rangle+\langle\vb{p}_aV_{12}^2\vb{p}_a\rangle\,.
\end{align*}
The first term of the right-hand side is {singular}, while the second term is regular.
In case of the molecular ions, the distribution $\rho_{12}$ is vanishingly small. As a consequence, the nucleus-nucleus divergence in $\langle V\vb{p}_a^2V\rangle$ cancels out algebraically. Therefore, the matrix element $\langle V\vb{p}_a^2V\rangle$ can be evaluated using the formula above within the cut-off regularization scheme.

All the modified matrix elements written on the right-hand side of equations in this section are finite and may be used in numerical calculations.

Yet another relation which connects the three basic singular distributions in the three-body case should be written
\begin{comment}
\begin{equation}\label{eq:va+rhoa}
\begin{array}{@{}l}\displaystyle
\frac{1}{2\mu_a}\Bigl(\left\langle \boldsymbol{\mathcal{E}}_a^2 \right\rangle+\left\langle V_a(4\pi\rho_a) \right\rangle\Bigr)
   -\left\langle V_a^3 \right\rangle =
	\frac{1}{2\mu_a}\left\langle \mathbf{p}_{a}V_a^2\mathbf{p}_{a} \right\rangle
	+\frac{1}{2\mu_b}\left\langle \mathbf{p}_{b}V_a^2\mathbf{p}_{b} \right\rangle
\\[3mm]\displaystyle\hspace{40mm}
	+\frac{1}{m}\left\langle \mathbf{p}_{a}V_a^2\mathbf{p}_{b} \right\rangle
	+\left\langle V_a^2(V_b+V_{12}) \right\rangle
	-E_0\left\langle V_a^2 \right\rangle.
\end{array}
\end{equation}
\end{comment}
\begin{align}\label{eq:va+rhoa}
	\langle V_a\rho_a\rangle - \langle \bm{\varepsilon}_a^2\rangle + 2\mu_a\langle V_a^3\rangle = \langle V_a\rho_a\rangle'\,,
\end{align}
where $\langle V_a\rho_a\rangle'$ is defined as follows
\begin{align}\label{eq:va+rhoa:reg}
	\langle V_a\rho_a\rangle' = \langle V_a^2\vb{p}_e^2\rangle'	-\langle \vb{p}_eV_a^2\vb{p}_e\rangle\,.
\end{align}

\subsection{Matrix elements $\langle\{\vb{p}_e^2,\rho_a\}\rangle$, $\langle\{\vb{p}_e^4,V_a\}\rangle$, $\langle\{\vb{p}_e^4,V_{12}\}\rangle$ and $\langle \vb{p}_e^2V_a\vb{p}_e^2\rangle$}

The finite part of the matrix element $\langle\{\vb{p}_e^2,\rho_a\}\rangle$ can be expressed using Eqs. (\ref{eq:SE_trans}) and \eqref{eq:va+rhoa}:
\begin{comment}
\begin{align}\label{eq:pes+rhoa}
\begin{array}{@{}l}\displaystyle
	\langle \{\vb{p}_e^2,\rho_a\}\rangle^{(m)}
	= -4\mu_a\langle V_a\rho_a\rangle_r +\mu_a\bigg\{
	4\left[E_0\langle\rho_a\rangle-\langle (V_b\!+\!V_{12})\rho_a\rangle\right]
\\[3mm]\displaystyle\hspace{53mm}
	-\frac{2}{m_a}\langle(\vb{p}_a^2\!-\!\vb{p}_e^2)\rho_a\rangle
	-\frac{2}{m_b}\langle \vb{p}_b^2\rho_a\rangle
	\bigg\}\,.
\end{array}
\end{align}
\end{comment}
\begin{align}\label{eq:pes+rhoa}
	\langle \{\vb{p}_e^2,\rho_a\}\rangle^{(m)} = -4\mu_a\left(\langle\bm{\varepsilon}_a^2\rangle_r-2\mu_a\langle{V}_a^3\rangle_r\right) + \langle\{\vb{p}_e^2,\rho_a\}\rangle'\,,
\end{align}
where
\begin{align}\label{eq:pes+rhoa:reg}
	\langle\{\vb{p}_e^2,\rho_a\}\rangle' &= -4\mu_a\langle V_a\rho_a\rangle'
	+\mu_a\bigg\{
	4\left[E_0\langle\rho_a\rangle-\langle (V_b\!+\!V_{12})\rho_a\rangle\right]
	\nonumber\\&
	-\frac{2}{m_a}\langle(\vb{p}_a^2\!-\!\vb{p}_e^2)\rho_a\rangle
	-\frac{2}{m_b}\langle \vb{p}_b^2\rho_a\rangle
	\bigg\}\,.
\end{align}

Next operator:
\begin{align}\label{eq:p4+va}
	\langle \{\vb{p}_e^4,V_a\}\rangle^{(m)} &= -4m_e\left(1\!-\!\frac{m_e}{m_a}\right)\left(\langle\bm{\varepsilon}_a^2\rangle_r-2\mu_a\langle{V}_a^3\rangle_r\right)
	\nonumber\\&\qquad
	+\langle \{\vb{p}_e^4,V_a\}\rangle'\,,
\end{align}
where
\begin{align}\label{eq:p4+va:reg}
\begin{array}{@{}l}\displaystyle
	\langle \{\vb{p}_e^4,V_a\}\rangle' = -4m_e\left(1\!-\!\frac{m_e}{m_a}\right)\langle V_a\vb{p}_e^2V_a\rangle'
	\\[3mm]\displaystyle\hspace{20mm}
	+4m_eE_0\left\langle V_a\left(\vb{p}_e^2-\frac{m_e}{m_a}\vb{p}_a^2-\frac{m_e}{m_b}\vb{p}_b^2\right)\right\rangle
	\\[3mm]\displaystyle\hspace{20mm}
	-4m_e\left[\langle\bm{\varepsilon}_a\bm{\varepsilon}_b\rangle+\langle V_a(V_b+V_{12})\vb{p}_e^2\rangle
	\right]
	\\[3mm]\displaystyle\hspace{20mm}
	+\frac{4m_e^2}{m_a}\Bigl[\pm\langle\bm{\varepsilon}_a\bm{\varepsilon}_{12}\rangle
	+\langle V_a^2(\vb{p}_a^2-\vb{p}_e^2)\rangle
	\\[3mm]\displaystyle\hspace{20mm}
	+\langle V_a(V_b+V_{12})\vb{p}_a^2\rangle\Bigr]
	+\frac{4m_e^2}{m_b}\langle V_aV\vb{p}_b^2\rangle\,.
\end{array}
\end{align}
Here higher-order terms in the expansion in $(m_e/m_a)^n$ are ignored. All matrix elements are regular.

The matrix element $\langle \vb{p}_e^2V_a\vb{p}_e^2\rangle$ can be expanded similarly:
\begin{align}\label{eq:pes+Va+pes}
	\langle \vb{p}_e^2V_a\vb{p}_e^2\rangle^{(m)} = 4\mu_a\left(1\!-\!\frac{m_e}{m_a}\right)\langle V_a^3\rangle_r + \langle\vb{p}_e^2V_a\vb{p}_e^2\rangle'\,,
\end{align}
where
\begin{align}\label{eq:pes+Va+pes:reg}
\begin{array}{@{}l}\displaystyle
	\langle\vb{p}_e^2V_a\vb{p}_e^2\rangle' =
	-2m_e\left(1\!-\!\frac{1}{m_a}\right)\langle V_a^2\vb{p}_e^2\rangle'
	\\[3mm]\displaystyle\hspace{10mm}
	\!+\!2m_eE_0\left\langle V_a\left(\vb{p}_e^2\!-\!\frac{m_e}{m_a}\vb{p}_a^2\!-\!\frac{m_e}{m_b}\vb{p}_b^2\right)\right\rangle
	\!+\!\frac{2m_e^2}{m_b}\langle V_aV\vb{p}_b^2\rangle
	\hspace*{-10mm}
	\\[3mm]\displaystyle\hspace{10mm}
	-2m_e\left[\langle V_a(V_b\!+\!V_{12})\vb{p}_e^2\rangle
	-\frac{m_e}{m_a}\langle V_a(V_b\!+\!V_{12})\vb{p}_a^2\rangle
	\right]
	\\[3mm]\displaystyle\hspace{10mm}
	+\frac{2m_e^2}{m_a}\langle V_a^2(\vb{p}_a^2-\vb{p}_e^2)\rangle.
\end{array}
\end{align}

It can be seen that the matrix elements $\langle\{\vb{p}_e^4,V_a\}\rangle^{(m)}$ and $\langle \vb{p}_e^2V_a\vb{p}_e^2\rangle^{(m)}$ are related by the following identity
\begin{align}\label{eq:p4v-pvp2}
	&\langle \{\vb{p}_e^4,V_a\}\rangle^{(m)} - 2 \langle \vb{p}_e^2V_a\vb{p}_e^2\rangle^{(m)} =
	4m_e\left(1\!-\!\frac{m_e}{m_a}\right)\langle \bm{\varepsilon}_a^2\rangle_r
	\nonumber\\&\qquad
	-4m_e\bigg[
	\langle \bm{\varepsilon}_a\bm{\varepsilon}_b\rangle
	\mp \frac{m_e}{m_a}\langle \bm{\varepsilon}_a\bm{\varepsilon}_{12}\rangle
	\bigg]
\end{align}

The regular matrix element $\langle\{\vb{p}_e^4,V_{12}\}\rangle$ can be calculated as follows:
\begin{align}\label{eq:p4+vr}
	\langle\{\vb{p}_e^4,V_{12}\}\rangle = 2\langle \vb{p}_e^2V_{12}\vb{p}_e^2\rangle\,.
\end{align}

\subsection{Matrix element $\langle U(E_0-H_0)U\rangle$}\label{sec:UEHU}

The matrix element $\langle U(E_0-H_0)U\rangle$ is expanded by inserting Eq. (\ref{eq:U-definition}):
\begin{align}\label{eq:UEHU}
	&\langle{U}(E_0\!-\!H_0)U\rangle
	=\lambda_1^2\langle{V}_1(E_0\!-\!H_0)V_1\rangle+\lambda_2^2\langle{V}_2(E_0\!-\!H_0)V_2\rangle
	\nonumber\\&\qquad
	+2\lambda_1\lambda_2\langle{V}_1(E_0\!-\!H_0)V_2\rangle\,,
\end{align}
where singular terms of the type $\langle V_a(E_0-H_0)V_a\rangle$ can be simplified:
\begin{align}\label{eq:va+EH+va}
	\langle{V}_a(E_0-H_0)V_a\rangle
	&=\langle{V}_a^2E_0\rangle-\langle{V}_a^2V\rangle-\frac{1}{2m_a}\langle{V}_a\vb{p}_a^2V_a\rangle
	\nonumber\\&\quad
	-\frac{1}{2m_b}\langle{V}_a\vb{p}_b^2V_a\rangle-\frac{1}{2m_e}\langle{V}_a\vb{p}_e^2V_a\rangle
	\nonumber\\
	&=-\frac{1}{2m_e}\left(1\!+\!\frac{m_e}{m_a}\right)\langle\bm{\varepsilon}_a^2\rangle_r\,.
\end{align}

As was shown, the singular terms $\langle\bm{\varepsilon}_a^2\rangle$ after summation within the nonrecoil contribution are cancelled out.

The matrix element $\langle V_1(E_0\!-\!H_0)V_2\rangle$ may be simplified by substituting the Hamiltonian~\eqref{eq:H0}:
\begin{align}
	\langle V_1(E_0-H_0)V_2\rangle
	& = \left\langle V_1\left(E_0-\frac{\vb{p}_e^2}{2m_e}-\frac{\vb{p}_1^2}{2m_1}-\frac{\vb{p}_2^2}{2m_2}
	-V\right)V_2\right\rangle\nonumber\\
	& = E_0\langle V_1V_2\rangle
	- \frac{1}{2m_e}\langle V_1\vb{p}_e^2V_2\rangle
	- \frac{1}{2m_1}\langle V_1V_2\vb{p}_1^2\rangle
	\nonumber\\&\quad
	- \frac{1}{2m_2}\langle V_1V_2\vb{p}_2^2\rangle
	- \langle V_1V_2V\rangle \nonumber\\
	& = E_0\langle V_1V_2\rangle
	-\frac{1}{2m_e}\langle \bm{\varepsilon}_1\bm{\varepsilon}_2\rangle
	-\left\langle V_1V_2H_0
	\right\rangle \nonumber\\
	& = -\frac{1}{2m_e}\langle \bm{\varepsilon}_1\bm{\varepsilon}_2\rangle\,.
\end{align}
The above derivation uses the identity \eqref{eq:va+pes+vb} and the fact that the wave function is a solution of the Schrödinger equation $(E_0\!-\!H_0)\psi_0=0$.

\section{Derivation of modified effective Hamiltonian} \label{appdeix:effec-hamiltonian-modified}

In this appendix, the singularity separation for the effective Hamiltonian $\delta H_1$-$\delta H_6$ is performed in detail, along with the second-order contribution $E_B$ and $E_R$.

It has been proved in many articles that the singularities of the $m\alpha^6$-order nonrecoil contributions are algebraically cancelled. Therefore, the regular parts of those divergent matrix elements are removed as well. On the other hand, the singularities of the $m\alpha^6(m/M)$-order recoil contributions are removed while its regular part should be kept in the final expression.

\subsection{expansion of $\delta H_1$}

Let us consider the effective Hamiltonian $\delta H_1$,
\begin{align}
	\delta H_1 = \frac{\vb{p}_e^6}{16m_e^5}\,.
\end{align}
Using Eq. (\ref{eq:SE_trans}), the matrix element $\langle \vb{p}_e^6\rangle$ can be expanded as follows:
\begin{align}\label{eq:p6-expansion}
	\langle \vb{p}_e^6\rangle^{(m)} &= 2m_eE_0\langle \vb{p}_e^4\rangle-m_e\langle\{\vb{p}_e^4,V_{12}\}\rangle
	\nonumber\\&\quad
	-\sum_am_e\left(\langle\{\vb{p}_e^4,V_a\}\rangle^{(m)}+\frac{1}{2m_a}\langle\{\vb{p}_e^4,\vb{p}_a^2\}\rangle^{(m)}\right)\,,
\end{align}
where the last term has the expansion as follows (ignoring higher-order recoil corrections)
\begin{align}
	\frac{1}{m_a}\langle\{\vb{p}_e^4,\vb{p}_a^2\}\rangle^{(m)}
	&=\frac{8m_e^2}{m_a}\bigg(E_0^2\langle \vb{p}_a^2\rangle-2E_0\langle V\vb{p}_a^2\rangle
	\nonumber\\&\quad
	+\langle V\vb{p}_a^2V\rangle^{(m)}\bigg)+O(1/m_{a,b}^2)\,.
\end{align}
The terms $\langle\{\vb{p}_e^4,V_a\}\rangle^{(m)}$ and $\langle V\vb{p}_a^2V\rangle^{(m)}$ can be taken from Eqs. (\ref{eq:p4+va}) and (\ref{eq:v+pas+v}) respectively.
The regular matrix elements $\langle\{\vb{p}_e^4,V_{12}\}\rangle$ can be calculated directly, see Eq. (\ref{eq:p4+vr}).

Inserting expressions of all matrix elements into $\langle \vb{p}^6_e\rangle$, we can obtain the explicit expansion for the modified effective Hamiltonian $\langle\delta H_1^{(m)}\rangle$:
\begin{align}\label{app:dH1_m}
\begin{array}{@{}l}\displaystyle
	\langle\delta H'_1\rangle
	=\frac{1}{16m_e^4}
	\Bigg\{
	2E_0\langle \vb{p}_e^4\rangle-\langle\{\vb{p}_e^4,V_{12}\}\rangle
\\[3mm]\displaystyle\hspace{20mm}
	-\sum_a
	\bigg[
	\langle\{\vb{p}_e^4,V_a\}\rangle'
\\[3mm]\displaystyle\hspace{20mm}
	+\frac{4m_e^2}{m_a}\left(E_0^2\langle \vb{p}_a^2\rangle-2E_0\langle V\vb{p}_a^2\rangle+\langle V\vb{p}_a^2V\rangle'\right)
	\bigg]
	\Bigg\}\,.
 \end{array}
\end{align}
The singular part corresponding to $\langle\delta H_1\rangle$ takes the form
\begin{align}
	S_1 = \sum_a\frac{1}{4m_e^3}\bigg[
		\left(1-\frac{2m_e}{m_a}\right)\langle \bm{\varepsilon}_a^2\rangle - 2m_e\left(1-\frac{3m_e}{m_a}\right)\langle V_a^3\rangle
	\bigg]\,.
\end{align}

\subsection{expansion of $\delta H_2$}

The next term of the effective Hamiltonian is $\delta H_2$,
\begin{align}
	\delta H_2 = \frac{1}{128m_e^4}[\vb{p}_e^2,[\vb{p}_e^2,V]] +
	\sum_a\frac{3}{64m_e^4}\{\vb{p}_e^2,\rho_a\}\,.
\end{align}

The expectation value for the first operator can be simplified as follows
\begin{align}\label{eq:pes+pes+V-expansion}
	\langle[\vb{p}_e^2,[\vb{p}_e^2,V]]\rangle^{(m)}&=\langle \{\vb{p}_e^4,V\}\rangle^{(m)}-2\langle \vb{p}_e^2V\vb{p}_e^2\rangle^{(m)}
	\nonumber\\&
	=\sum_a \bigg(\langle\{\vb{p}_e^4,V_a\}\rangle^{(m)}-2\langle \vb{p}_e^2V_a\vb{p}_e^2\rangle^{(m)}\bigg)
	\nonumber\\&
	=-4m_e\sum_a\bigg[
	\langle\bm{\varepsilon}_a\bm{\varepsilon}_b\rangle\mp\frac{m_e}{m_a}\langle\bm{\varepsilon}_a\bm{\varepsilon}_{12}\rangle
	\nonumber\\&\quad
	+\left(1-\frac{m_e}{m_a}\right)\langle\bm{\varepsilon}_a^2\rangle_r
	\bigg]\,.
\end{align}
Here the identity of Eq. \eqref{eq:p4v-pvp2} is adopted to get the final result.

The second term of $\delta H_2$ is proportional to $\{\vb{p}_e^2,\rho_a\}$ and can be treated as in Eq. (\ref{eq:pes+rhoa}).

Combining the above matrix elements, one can obtained the modified $\delta H_2$ as
\begin{align}\label{app:dH2_m}
	\langle \delta H'_2\rangle
	=\sum_a\bigg[&-\frac{1}{32m_e^3}
	\bigg(
	\langle\bm{\varepsilon}_a\bm{\varepsilon}_b\rangle\mp\frac{m_e}{m_a}\langle\bm{\varepsilon}_a\bm{\varepsilon}_{12}\rangle
	\bigg)
	\nonumber\\&\quad
	+\frac{3}{64m_e^4}\langle\{p_e^2,\rho_a\}\rangle'
	\bigg]\,,
\end{align}
And the singular part is
\begin{align}
	S_2 =\sum_a
	\bigg[&
		-\frac{7}{32m_e^3}\left(1-\frac{m_e}{m_a}\right)\langle\bm{\varepsilon}_a^2\rangle
		\nonumber\\&
		+\frac{3}{8m_e^2}\left(1-\frac{2m_e}{m_a}\right)\langle V_a^3\rangle
	\bigg]\,.
\end{align}

\subsection{expansion of $\delta H_3$}

The effective Hamiltonian $\delta H_3$ has the form of recoil contribution, but is identified as part the nonrecoil contribution,
\begin{align}
	\delta H_3 = \sum_a-\frac{1}{32m_e^3m_a}[\vb{p}_e^2,[\vb{p}_a^2,V_a]]\,.
\end{align}
The matrix element $\langle [\vb{p}_e^2,[\vb{p}_a^2,V_a]] \rangle/m_a$ is transformed in a similar way as in Eq. (\ref{eq:pes+pes+V-expansion}):
\begin{align}
	\frac{1}{m_a}\langle [\vb{p}_e^2,[\vb{p}_a^2,V_a]] \rangle^{(m)}
	&=\frac{2m_e}{m_a}\langle [[\vb{p}_a^2,V_a],V]\rangle^{(m)}
	\nonumber\\&
	=\frac{4m_e}{m_a}
	\left(
	\langle VV_a\vb{p}_a^2\rangle^{(m)} - \langle V_a\vb{p}_a^2V\rangle^{(m)}
	\right)
	\nonumber\\
	&=\frac{4m_e}{m_a}
	\bigg(
	\langle V_a^2\vb{p}_a^2\rangle^{(m)} - \langle V_a\vb{p}_a^2V_a\rangle^{(m)}
	\nonumber\\&\quad
	+\langle V_aV_{12}\vb{p}_a^2\rangle - \langle V_a\vb{p}_a^2V_{12}\rangle
	\bigg)
	\nonumber\\&
	=-\frac{4m_e}{m_a}\left(\langle \bm{\varepsilon}_a^2\rangle_r \pm\langle\bm{\varepsilon}_a\bm{\varepsilon}_{12}\rangle \right)\,.
\end{align}

Therefore, one gets the modified $\delta H_3$ in the form
\begin{align}
	\langle \delta H'_3\rangle
	=\sum_a\mp\frac{1}{8m_e^2m_a}\langle\bm{\varepsilon}_a\bm{\varepsilon}_{12}\rangle\,,
\end{align}
while the singular part is separated as
\begin{align}
	S_3 =\sum_a \frac{1}{8m_e^2m_a}\langle\bm{\varepsilon}_a^2\rangle \,.
\end{align}

\subsection{expansion of $\delta H_4$}\label{appendix:dH4}

Now we turn to the effective Hamiltonian $\delta H_4$, which has been transferred as Eq. (\ref{eq:dH4_expansion_1st}):
\begin{align}
	\langle\delta H_4\rangle = -\frac{E_0}{m_e}  \langle H_R\rangle
	-\sum_a\frac{z_ez_a}{2m_e^2m_a}\langle \{V,~p_e^i\mathcal{W}^{ij}_ap_a^j \}\rangle + O(1/m_{a,b}^2)\,.
\end{align}
the divergent matrix element $\langle \{V,~p_e^i\mathcal{W}^{ij}_ap_a^j \}\rangle$ is expanded as Eq. (\ref{eq:dH4_vwp}) by acting momentum operators on the Coulomb potential,
\begin{align}
	\label{eq:dH4_vwp_app}
	\langle \{V, p_e^i\mathcal{W}^{ij}_ap_a^j\} \rangle
	&= 2\langle p_e^i\mathcal{W}^{ij}_aVp_a^j\rangle
	- \langle \imath\varepsilon_a^i\mathcal{W}_a^{ij}p_a^j\rangle
	- \langle p_e^i\mathcal{W}_a^{ij}\imath\varepsilon_a^j\rangle\,.
\end{align}

Let us consider the second term in Eq. (\ref{eq:dH4_vwp_app}). Adopting the identity $\vb{p}_e+\vb{p}_1+\vb{p}_2=0$, it can be split into two terms:
\begin{align}\label{eq:epsa+wa+pa}
	z_ez_a\langle \imath\varepsilon_a^i\mathcal{W}_a^{ij}p_a^j\rangle
	=\langle V_a\imath\bm{\varepsilon}_a\vb{p}_a\rangle
	=-\langle V_a\imath\bm{\varepsilon}_a\vb{p}_e\rangle
	-\langle V_a\imath\bm{\varepsilon}_a\vb{p}_b\rangle\,.
\end{align}
One can see that the first term is singular and the second term is regular.
Meanwhile, it has been numerical {verified} that matrix elements $\langle V_a\imath\bm{\varepsilon}_a\vb{p}_b\rangle$ are smaller than 10$^{-20}$ and can be removed from final expressions safely.
%small enough to be ignored.
We find that the last matrix element in Eq. (\ref{eq:dH4_vwp_app}) is singular,
\begin{align}\label{eq:pe+wa+epsa}
	z_ez_a\langle p_e^i\mathcal{W}^{ij}_a\imath\varepsilon_a^j\rangle
	=\langle \vb{p}_e\imath\bm{\varepsilon}_aV_a\rangle\,.
\end{align}

Inserting all matrix elements into Eq. (\ref{eq:dH4_vwp}) yields the modified $\langle \delta H_4^{(m)}\rangle$ while divergent matrix elements are removed
\begin{align}\label{eq:app-H4m}
	\langle \delta H'_4\rangle	
	=-\frac{E_0}{m_e}\langle H_R\rangle - \sum_a \frac{1}{2m_e^2m_a}
	z_ez_a\langle 2p_e^i\mathcal{W}^{ij}_aVp_a^j\rangle
	\,,
\end{align}
and the singularity of $\delta H_4$ is then has a form:
\begin{align}\label{eq:app-S4}
	S_4 &=\sum_a\frac{1}{2m_e^2m_a}(\langle \vb{p}_e\imath\bm{\varepsilon}_aV_a\rangle-\langle V_a\imath\bm{\varepsilon}_a\vb{p}_e\rangle)
	\nonumber\\
	&= \sum_a\frac{1}{2m_e^2m_a}\left(\langle V_a\rho_a\rangle-\langle\bm{\varepsilon}_a^2\rangle\right)\,.
\end{align}
In the above, we adopt identities $\langle \vb{p}_e\imath\bm{\varepsilon}_aV_a\rangle+\langle V_a\imath\bm{\varepsilon}_a\vb{p}_e\rangle\equiv0$ and $2\langle V_a\imath\bm{\varepsilon}_a\bm{p}_e\rangle =  \langle\varepsilon_a^2\rangle-\langle V_a\rho_a\rangle $ (see Eq. (19) of Ref. \cite{Korobov2025reg}). 

\subsection{expansion of $\delta H_5$}

The modified effective Hamiltonian $\langle \delta H_5^{(m)}\rangle$ can be obtained easily,
\begin{align}\label{eq:app-H5m}
	\langle \delta H'_5\rangle = \sum_a\frac{1}{8m_e^2m_a}
	\bigg[
	\langle
	\bm{p}_eV_a^2\bm{p}_e
	\rangle
	+3\langle
	(\bm{p}_e\cdot\hat{r}_a)V_a^2(\hat{r}_a\cdot\bm{p}_e)
	\rangle
	\bigg]
	\,,
\end{align}
while the singularity of this term is
\begin{align}\label{eq:app-S5}
	S_5 = \sum_a\frac{1}{4m_e^2m_a}\langle\bm{\varepsilon}_a^2\rangle\,.
\end{align}

\subsection{expansion of $\delta H_6$}\label{sec:dH6-app}

The effective Hamiltonian $\delta H_6$ has a formula of:
\begin{align}
	\delta H_6=\sum_a-\frac{z_ez_a}{m_em_a}
	\bigg\{ &
	[p_e^i,V]\mathcal{X}^{ij}_a[V,p_a^j]
	\nonumber\\&
	+p_e^i\left[\mathcal{X}^{ij}_a,\frac{\vb{p}_e^2}{2m_e}\right][V,p_a^j]
	\bigg\}\,.
\end{align}

The first matrix element is singular and can be expanded by performing $[p_e^i,V]$ and $[V,p_a^j]$:
\begin{align}\label{eq:dH6_1st-term}
	&z_ez_a\langle[p_e^i,V]\mathcal{X}^{ij}_a[V,p_a^j]\rangle
	\nonumber\\
	&=\!z_ez_a
	\bigg\langle
	\imath(\varepsilon_a\!+\!\varepsilon_b)^i\mathcal{X}^{ij}_a\imath\varepsilon_a^j
	\bigg\rangle
	\!\mp\! z_ez_a\bigg\langle
	\imath(\varepsilon_a\!+\!\varepsilon_b)^i\mathcal{X}^{ij}_a\imath\bm{\varepsilon}_{12}^j
	\bigg\rangle
	\nonumber\\
	&=\frac{1}{4}\langle V_a^3\rangle
	+\frac{z_ez_a}{4}\langle
	r_a\bm{\varepsilon}_a\bm{\varepsilon}_b
	\rangle
	\mp z_ez_a\bigg\langle
	\imath(\varepsilon_a+\varepsilon_b)^i\mathcal{X}^{ij}_a\imath\bm{\varepsilon}_{12}^j
	\bigg\rangle\,.
\end{align}

Let us evaluate the second term of $\delta H_6$.
Firstly, we act by the operator $\vb{p}_e^2$ on $\mathcal{X}^{ij}_a$ to expand the commutator
\begin{align}
	\left[\mathcal{X}^{ij}_a,\frac{\vb{p}_e^2}{2m_e}\right]&
	=\frac{1}{m_e}\bigg[\left(\partial^k\mathcal{X}^{ij}_a\right)\partial^k
	+\frac{1}{2}\left(\partial^k\partial^k\mathcal{X}^{ij}_a\right)\bigg]\,,
\end{align}
where
\begin{align}
	\partial^k\mathcal{X}^{ij}
	=\frac{1}{8}\left(\delta^{ki}\frac{r^j}{r}+\delta^{kj}\frac{r^i}{r}-\frac{r^ir^jr^k}{r^3}-3\delta^{ij}\frac{r^k}{r}\right)\,,
\end{align}
and
\begin{align}
	\partial^k\partial^k\mathcal{X}^{ij}
	=-\frac{1}{2r}\left(\delta^{ij}+\frac{r^ir^j}{r^2}\right)\,.
\end{align}
After that, the second term of $\delta H_6$ can be split into two terms.
The first term is
\begin{align}
	&z_ez_a\bigg\langle p_e^i  \left(\partial_e^k\mathcal{X}^{ij}_a\right)\partial_e^k\imath\varepsilon_a^j\bigg\rangle
	=\imath(z_ez_a)^2\bigg\langle
	p_e^i\left(\partial_e^k\mathcal{X}^{ij}_a\right)
	\bigg[\frac{\delta^{kj}}{3}4\pi\delta(\vb{r}_a)
	\nonumber\\&\quad
	+\frac{1}{r_a^3}\left(\delta^{kj}-3\frac{r_a^jr_a^k}{r_a^2}\right)\bigg]
	\bigg\rangle
%	\nonumber\\&\quad
	+z_ez_a\bigg\langle
	p_e\left(\partial_e^k\mathcal{X}^{ij}_a\right)\imath\varepsilon_a^j\partial_e^k
	\bigg\rangle
	\nonumber\\&
	=\frac{3}{4}\langle\vb{p}_e\imath\bm{\varepsilon}_aV_a\rangle
	-\frac{1}{8}\bigg[\langle \vb{p}_eV_a^2\vb{p}_e\rangle-3\left\langle (\vb{p}_e\cdot\vb{r}_a)\frac{V_a^2}{r_a^2}(\vb{r}_a\cdot\vb{p}_e)\right\rangle\bigg]\,.
\end{align}
The second one is
\begin{align}
	\frac{z_ez_a}{2}\bigg\langle
	p_e^i(\partial_e^k\partial_e^k\mathcal{X}^{ij}_a)\imath\varepsilon_a^j
	\bigg\rangle
	=-\frac{1}{2}\langle\vb{p}_e\imath\bm{\varepsilon}_aV_a\rangle\,.
\end{align}
Thus, the second term of $\delta H_6$ can be evaluated as
\begin{align}\label{eq:dH6_2nd-term}
	&z_ez_a\bigg\langle p_e^i\left[\mathcal{X}^{ij}_a,\frac{\vb{p}_e^2}{2m_e}\right][V,p_a^j]
	\bigg\rangle
	=\frac{1}{8m_e}\left[\langle V_a\rho_a\rangle - \langle \bm{\varepsilon}_a^2 \rangle\right]
	\nonumber\\&\qquad
	-\frac{1}{8m_e}
	\bigg[
	\left\langle \vb{p}_eV_a^2\vb{p}_e\right\rangle
	-3\left\langle (\vb{p}_e\cdot\vb{r}_a)\!\frac{V_a^2}{r_a^2}\!(\vb{r}_a\cdot\vb{p}_e)\right\rangle
	\bigg]\,.
\end{align}

Now we can insert Eqs. (\ref{eq:dH6_1st-term}) and (\ref{eq:dH6_2nd-term}) into $\langle\delta H_6\rangle$ to obtained the expectation value of the modified effective Hamiltonian
\begin{align}\label{eq:app-H6m}
	\langle \delta H'_6\rangle
	&=\sum_a\frac{1}{8m_e^2m_a}
	\bigg\{
	\left\langle \vb{p}_eV_a^2\vb{p}_e\right\rangle
	-3\left\langle (\vb{p}_e\cdot\vb{r}_a)\!\frac{V_a^2}{r_a^2}\!(\vb{r}_a\cdot\vb{p}_e)\right\rangle
	\nonumber\\&
	-2z_ez_am_e\langle r_a\bm{\varepsilon}_a\bm{\varepsilon}_b\rangle
	\pm 8z_ez_am_e\bigg\langle
	\imath(\varepsilon_a+\varepsilon_b)^i\mathcal{X}^{ij}_a\imath\bm{\varepsilon}_{12}^j
	\bigg\rangle
	\bigg\}
	\,,
\end{align}
while its singular part has the form of
\begin{align}\label{eq:app-S6}
	S_6 = \sum_a\frac{1}{8m_e^2m_a}\left[\langle\bm{\varepsilon}_a^2\rangle-2m_e\langle V_a^3\rangle-\langle V_a\rho_a\rangle\right]\,.
\end{align}

\subsection{evaluation of $H_B^\text{1st}$}

This term generated from the transferred second-order correction $E_B$ ( see Eq. \eqref{eq:HB-1st})
\begin{align}\label{eq:HB-app}
	H_B^\text{1st} = \{H_B,U\} - 2\langle{H_B}\rangle U - U(E_0-H_0)U\,.
\end{align}
In the evaluation, its first and third terms contain singularities.

The first term of $H_B^\text{1st}$ is expanded by inserting $U=\sum_a\lambda_aV_a$ \eqref{eq:U-definition} and $H_B$ \eqref{eq:ma4-HB}
\begin{align}
	\{H_B,U\}=\sum_a-\frac{\lambda_a}{8m_e^3}\bigg[\{\vb{p}_e^4,V_a\}+m_e\{V_a,\rho_a+\rho_b\}\bigg]\,.
\end{align}
The matrix elements $\langle \{\vb{p}_e^4,V_a\}\rangle^{(m)}$ and $\langle V_a\rho_a\rangle^{(m)}$ can be taken from Eqs. (\ref{eq:p4+va}) and (\ref{eq:va+rhoa}), respectively.

The third term $\langle U(E_0-H_0)U\rangle$ in \eqref{eq:HB-app} has been considered in Sec.~\ref{sec:UEHU}.

Therefore, the regularized effective Hamiltonian of $H_B^\text{1st}$ is
\begin{align}\label{eq:app-HB-1stm}
\begin{array}{@{}l}\displaystyle
	\langle {H'}_B^\text{1st}\rangle = -2\langle H_B\rangle\langle U\rangle
	+\frac{\lambda_1\lambda_2}{m_e}\langle\bm{\varepsilon}_1\bm{\varepsilon}_2\rangle
\\[2mm]\displaystyle
	-\sum_a\frac{\lambda_a}{8m_e^3}\left(\langle\{\vb{p}_e^4,V_a\}\rangle'+2m_e\langle V_a\rho_b\rangle
	+2m_e\langle V_a\rho_a\rangle'\right)\,,
\end{array}
\end{align}
while the corresponding singular term may be written as
\begin{align}\label{eq:app-SB}
	S_B = \sum_a
	\bigg[
		-\frac{1}{32m_e^3}\left(1-\frac{5m_e}{m_a}\right)\langle\bm{\varepsilon}_a^2\rangle
		+\frac{1}{8m_e^2}\left(1-\frac{6m_e}{m_a}\right)\langle V_a^3\rangle
	\bigg]\,.
\end{align}

\subsection{evaluation of $H_R^\text{1st}$}\label{app:H_R-1st}

This term arises from the contribution of the second-order correction $E_R$ (see Eq. (\ref{eq:ER_expansion})):
\begin{align}
	H_R^\text{1st}=\{H_R,U\}-2\langle H_R\rangle U\,.
\end{align}
Taking into account the expression of $H_R$, Eq. (\ref{eq:ma4-HR}), the first term of $H_R^\text{1st}$ is expressed as
\begin{align}
	\langle \{H_R,U\}\rangle
	=\sum_a\sum_b -\lambda_b\frac{z_ez_a}{m_em_a} \langle \{ p_e^i\mathcal{W}_a^{ij}p_a^j,V_b\}\rangle\,.
\end{align}
In the above formula, one can see that the cross terms with $a\neq b$ are regular, while the terms with $a=b$ are singular.
The regular matrix element can be evaluated as
\begin{align}
	z_ez_a\langle\{p_e^i\mathcal{W}_a^{ij}p_a^j,V_b\}\rangle_{b\neq a}
	&=2z_ez_a\left\langle p_e^i\mathcal{W}_a^{ij}V_bp_a^j\rangle
	-\langle\imath\bm{\varepsilon}^i_b\mathcal{W}_a^{ij}p_a^j\right\rangle\,.
\end{align}
where the second term on the right-hand side is negligibly small and can be removed from final expressions safely.
The divergent matrix element $\langle \{V_a,p_e^i\mathcal{W}_a^{ij}p_a^j\}\rangle$ can be evaluated as
\begin{align}
	&z_ez_a\langle\{p_e^i\mathcal{W}_a^{ij}p_a^j,V_a\}\rangle_{b=a}
	\!=\!2z_ez_a\langle p_e^i\mathcal{W}_a^{ij}V_ap_a^j\rangle
	\!-\!\langle\imath\varepsilon_a^i\mathcal{W}_a^{ij}p_a^j\rangle
	\nonumber\\&\qquad
	\!-\!\langle p_e^i\mathcal{W}_a^{ij}\imath\varepsilon_a^j\rangle
	\nonumber\\
	&=	2z_ez_a\langle p_e^i\mathcal{W}_a^{ij}V_ap_a^j\rangle
	+\langle V_a\imath\bm{\varepsilon}_a\vb{p}_e\rangle
	+\langle V_a\imath\bm{\varepsilon}_a\vb{p}_b\rangle
	-\langle \vb{p}_e\imath\bm{\varepsilon}_aV_a\rangle\,,
\end{align}
where the identities of Eqs. (\ref{eq:epsa+wa+pa}) and (\ref{eq:pe+wa+epsa}) are adopted.
Again, matrix element $\langle V_a\imath\bm{\varepsilon}_a\vb{p}_b\rangle$ is small enough to ignore.

Therefore, the expectation value for the regularized effective  Hamiltonian is obtained by summing up all results
and ignoring higher order recoil contributions
\begin{align}\label{eq:app-HR-1stm}
	\langle {H'}_R^\text{1st}\rangle
	&=
	\sum_{b\neq a}
		-\frac{2z_ez_a}{m_e^2m_a}\langle p_e^i\mathcal{W}_a^{ij}Up_a^j\rangle
	-2\langle H_R\rangle \langle U\rangle
	\nonumber\\&\quad
	+O(1/m_{p,d}^2)\,,
\end{align}
while the singularity of $E_R$ can be isolated as
\begin{align}\label{eq:app-SR}
	S_R&=\sum_a -\frac{\lambda_a}{m_em_a}\left[\langle V_a\imath\bm{\varepsilon}_a\vb{p}_e\rangle-\langle \vb{p}_e\imath\bm{\varepsilon}_aV_a\rangle\right]
		\nonumber\\&
	=\sum_a \frac{1}{4m_e^2m_a}\left[\langle\bm{\varepsilon}_a^2\rangle-\langle V_a\rho_a\rangle\right]+O(1/m_{p,d}^2)\,.
\end{align}

\section{$m\alpha^6(m/M)$-order recoil corrections in hydrogen atom}\label{app:H-ma6-rec}

Using the derivations of Sec.~\ref{appdeix:effec-hamiltonian-modified}\, for the case of the hydrogen atom, the effective Hamiltonian for the relativistic recoil corrections of order $m\alpha^6(m/M)$ is obtained in the form
\begin{align}\label{app:D1}
	H^{(6m)}_\text{rec}(\text{H}) & = \frac{\bm{\varepsilon}^2}{8M}-\frac{V^3}{4M}+\frac{V\rho}{8M}
	+\frac{(z_ez_N)^2}{4M}\vb{p}\frac{1}{r^2}\vb{p}
	\nonumber\\&\quad
	+\frac{(z_ez_N)^2}{4M}\left(\vb{p}\frac{1}{r^2}\vb{p}+\frac{(\vb{p}\vb{r})(\vb{r}\vb{p})}{r^4}\right)
	\nonumber\\&\quad
	-E_0H_R-2\langle H_R\rangle  U
	+C_r\frac{z_N^3}{M}\pi\delta(\vb{r})\,.
\end{align}
where the operators $\bm{\varepsilon}^2$, $V^3$, and $V\rho$ are considered to be finite and obtained by means of the coordinate cutoff regularization. The matrix element $\langle V\rho\rangle$ is nonzero and proportional to $\langle\delta(\mathbf{r})$$\rangle$. The coefficient $C_r$ should be defined from comparison of $\langle H_\text{rec}^{(6m)}\rangle$ from Eq.~\eqref{app:D1} with the known result for the hydrogen $nS$ states.

The mean values of matrix elements for the \emph{nS} states of hydrogen case we take from Eq. (B1) of Ref. \cite{Korobov2025reg} and Eqs. (23-24) of Ref. \cite{Korobov2025rec}.
Thus the expectation value of $H_\text{rec}^{(6m)}$ can be expressed as
\begin{align}
	E_\text{rec}^{(6m)} = \frac{m^2z_N^6}{Mn^3}\left(-\frac{23}{6}-\frac{3}{n}+\frac{35}{6n^2}-\frac{2}{n^3}\right)\,.
\end{align}
This result should be compared with the analytical formula for the total $m\alpha^6$-order correction for the two-body system \cite{Pachucki1997ma6}. This allows to find the {unknown} coefficient $C_r$:
\begin{align}
	C_r = 4\ln2-2\,,
\end{align}
and the last contribution in Eq.~\eqref{app:D1} includes the high-energy part \eqref{eq:H_H} as well.

%\bibliography{myrefs.bib}

\end{document}